\documentclass[twocolumn]{aastex631}

\usepackage{booktabs}
\usepackage{appendix}
\usepackage{tabularx}
\usepackage{booktabs,makecell}
\usepackage{nameref}
\usepackage{array}
\usepackage[flushleft]{threeparttable}

\usepackage{lipsum}                     
\usepackage{xargs}     
\usepackage{float}

\usepackage{xcolor}
\usepackage{amsmath}

\begin{document}

\title{A Systematic Search for Main-Sequence Dipper Stars Using the Zwicky Transient Facility}

\author[0000-0003-0484-3331]{Anastasios Tzanidakis}\thanks{Corresponding author: A. Tzanidakis, \href{mailto:atzanida@uw.edu}{atzanida@uw.edu}}
\affiliation{Department of Astronomy and the DiRAC Institute, University of Washington, 3910 15th Avenue NE, Seattle, WA 98195, USA}

\author[0000-0002-0637-835X]{James R. A. Davenport}
\affiliation{Department of Astronomy and the DiRAC Institute, University of Washington, 3910 15th Avenue NE, Seattle, WA 98195, USA}

\author[0000-0003-3287-5250]{Neven Caplar}
\affiliation{Department of Astronomy and the DiRAC Institute, University of Washington, 3910 15th Avenue NE, Seattle, WA 98195, USA}

\author[0000-0001-8018-5348]{Eric C. Bellm}
\affiliation{Department of Astronomy and the DiRAC Institute, University of Washington, 3910 15th Avenue NE, Seattle, WA 98195, USA}

\author[0009-0003-1791-8707]{Wilson Beebe}
\affiliation{Department of Astronomy and the DiRAC Institute, University of Washington, 3910 15th Avenue NE, Seattle, WA 98195, USA}

\author[0009-0009-7822-7110]{Doug Branton}
\affiliation{Department of Astronomy and the DiRAC Institute, University of Washington, 3910 15th Avenue NE, Seattle, WA 98195, USA}

\author[0009-0007-9870-9032]{Sandro Campos}
\affiliation{McWilliams Center for Cosmology and Astrophysics, Department of Physics, Carnegie Mellon University, Pittsburgh, PA 15213, USA}

\author[0000-0001-5576-8189]{Andrew J. Connolly}
\affiliation{Department of Astronomy and the DiRAC Institute, University of Washington, 3910 15th Avenue NE, Seattle, WA 98195, USA}
\affiliation{eScience Institute, University of Washington, Seattle, WA 98195, USA}

\author[0000-0002-1074-2900]{Melissa DeLucchi}
\affiliation{McWilliams Center for Cosmology and Astrophysics, Department of Physics, Carnegie Mellon University, Pittsburgh, PA 15213, USA}

\author[0000-0001-7179-7406]{Konstantin Malanchev}
\affiliation{McWilliams Center for Cosmology and Astrophysics, Department of Physics, Carnegie Mellon University, Pittsburgh, PA 15213, USA}

\author[0009-0005-8764-2608]{Sean McGuire}
\affiliation{McWilliams Center for Cosmology and Astrophysics, Department of Physics, Carnegie Mellon University, Pittsburgh, PA 15213, USA}

\begin{abstract}
Main-sequence dipper stars, characterized by irregular and often aperiodic luminosity dimming events, offer a unique opportunity to explore the variability of circumstellar material and its potential links to planet formation, debris disks, and broadly star-planet interactions. The advent of all-sky time-domain surveys has enabled the rapid discovery of these unique systems. We present the results of a large systematic search for main-sequence dipper stars, conducted across a sample of 63 million FGK main-sequence stars using data from Gaia eDR3 and the Zwicky Transient Facility (ZTF) survey. Using a novel light curve scoring algorithm and a scalable workflow tailored for analyzing millions of light curves, we have identified 81 new dipper star candidates. Our sample reveals a diverse phenomenology of light curve dimming shapes, such as skewed and symmetric dimmings with timescales spanning days to years, some of which closely resemble exaggerated versions of KIC 8462852. Our sample reveals no clear periodicity patterns sensitive to ZTF in many of these dippers and no infrared excess or irregular variability. Using archival data collated for this study, we thoroughly investigate several classification scenarios and hypothesize that the mechanisms of such dimming events are either driven by circumstellar clumps or occultations by stellar/sub-stellar companions with disks. Our study marks a significant step forward in understanding main-sequence dipper stars.
\end{abstract}


\newpage

\section{Introduction} \label{sec:intro}

The era of synoptic time-domain surveys has revolutionized the taxonomy of variable stars. Among these variables is an emerging new class of main-sequence stars, dubbed as ``main-sequence dippers'' that can exhibit non-periodic variations in their luminosity and whose origin is still largely debated in the community. The most well-known system is KIC 8462852 (hereafter ``Boyajian's star'') \citep{2016MNRAS.457.3988B}, a seemingly ordinary F3V dwarf star that underwent irregular photometric deep dimming events, including a long-term secular dimming \citep{2016ApJ...830L..39M} during the primary Kepler survey \citep{Borucki2010}. The four main observed features to be interpreted of the Boyajian star system are the aperiodic dimming events during the Kepler mission, the transit shape of the dimming events, a secular dimming on the timescale of years to centuries \citep{2016ApJ...822L..34S, 2016ApJ...830L..39M}, and the lack of infrared excess indicating the presence of warm dust \citep{2015ApJ...814L..15M}. \citet{2017A&A...608A.132K} showed that two dimming events in the Boyajian star system showed similarities, occurring 928.25 days apart, suggesting a potential repeated transit of a large ring system or a string of exocomets. Despite extensive follow-up studies, pinpointing any periodicity in the dimming events remains unconfirmed \citep{Boyajian2018}, but follow-up multiband photometry shows the dimming events show the reddening of non-grey extinction of sub-micron dust. The interpretation of the observable dimming events and lack of infrared excess continues to spark debate in the literature, with possible explanations including transiting exocomets \citep{2016MNRAS.457.3988B}, comet swarms \citep{2016ApJ...833...78M, 2016ApJ...819L..34B}, large ringed planets with Trojan asteroid swarms \citep{2018MNRAS.473L..21B}, the consumption of a planet \citep{2017MNRAS.468.4399M}, the tidal disruption of orphan exomoons \citep{2019MNRAS.489.5119M}, or the presence of a potential technosignature of a transiting megastructure \citep{2016ApJ...816...17W}. A comprehensive review of these competing theories, including an analysis of their advantages and disadvantages, is provided by \citet{TabbySolutions} and references therein. 

More recently, ASASSN-21qj \citep{2021ATel14879....1R}, a G dwarf, underwent a dramatic dimming event that was later attributed to either the collision of two planetesimals producing a debris field \citep{2023Natur.622..251K} or the breakup of exocometary bodies \citep{2023ApJ...954..140M}. Unlike the Boyajian star, ASASSN-21qj exhibited an optical dimming event while the infrared data from the \textit{Wide-field Infrared Survey Explorer} (WISE; \citet{WISE_citation, wise_doi}) showed a brightening that lagged behind the optical dimming. The infrared excess observed was constrained to originate between 2 and 16 AU from the star, suggesting that the debris was warm enough ($\sim$1000 Kelvin) to emit significant infrared radiation that is hypothesized to be involved in the collision of two Earth-size planets \citep{2023Natur.622..251K}. Findings from \citet{2023ApJ...954..140M} suggest the occulting material consists of submicron-sized dust grains at much smaller semimajor axes of $\sim$0.2 AU, with a minimum total circumstellar dust mass of 10$^{-6}$M$_{\Terra}$, alluding to a different origin and timescale for the dust production. No additional large amplitude dimming events have been reported from this system to date. Although both ASASSN-21qj and the Boyajian star exhibit aperiodic and complex dimming events, ASASSN-21qj shows an infrared excess from warm dust, indicating more recent and possibly close-in material. In contrast, no infrared excess has been detected around the Boyajian star, suggesting that the dust production mechanisms differ in timescale, mass, orbital geometry, and other parameters. 

Another class of poorly understood variable stars exhibiting similar dimming behavior involves rare occultations by large and or opaque disks, typically associated with both young and evolved stars. \citet{2012AJ....143...72M} suggests that a survey of young ($\sim$10 Myr) post-accretion pre-main sequence (PMS) stars should yield a handful of deep eclipses caused by circumplanetary disks and disks surrounding low-mass companion stars, due to the high frequency of binary systems in young clusters, many of which can likely host circumplanetary or circumsecondary disks (e.g., \citet{2007arXiv0705.3258P}). \citet{2012AJ....143...72M} presented one such candidate, 1SWASP J140747.93-394542.6 (hereafter J1407) which was a nearby young ($\sim$16Myr) weak-lined T Tauri star that underwent a series of complex eclipses that lasted for two months and is believed to be caused by an unseen secondary companion, J1407b, with a giant ring system with gaps with an unconfirmed orbital period \citep{2015MNRAS.446..411K, 2015ApJ...800..126K}. In contrast, PDS 110 a $\sim$10 Myr star with large infrared excess in the Ori OB1a association was found to be eclipsed by a low-mass ($<$70 M$_{\text{Jup}}$) companion, PDS 110b, on an $\sim$ 800-day orbital period, with a circum-secondary disk around a low-mass planet or brown dwarf of diameter $\sim$0.3 AU \citep{2017MNRAS.471..740O}. The shared characteristics of J1407b and PDS 110b are both located in regions of ongoing star formation, suggesting stellar youth. In more extreme cases involving more evolved primary stars, systems such as $\epsilon$ Aurigae and other analog systems \citep{2025arXiv250703080F, 2025arXiv250705367Z, 2024A&A...688L..11P, 2023ApJ...955...69T, 2019MNRAS.482.5000S, 2016AJ....151..123R, 2010Natur.464..870K} are believed to result from occultations by large opaque disks on extremely long orbital periods. The dipper phenomenon is more common among Young Stellar Objects (YSO) that have been found to exhibit periodic, quasiperiodic, or aperiodic dimming events from timescales of hours to years \citep{2023ASPC..534..355F, 2014AJ....147...82C}, attributed to line-of-sight effects of magnetospheric accretion disk columns and dust around PMS stars \citep{Bodman_Dippers_YSO}. Large-scale monitoring campaigns, such as the CSI-2265 project, have shown that over 20$\%$ of disk-bearing stars exhibit discrete fading events proposed to be circumstellar obscurations or changes in the inner disk. Similar detection rates of YSO dipper stars have been reported among young stellar associations in high-precision photometric missions, Kepler/K2 and the Transiting Exoplanet Survey Satellite \citep{Ricker}, \citep{TESS_dipper, Hedghes18}.

Resolving the main-sequence dipper phenomenon requires careful systematic investigations into analog systems and shared properties. Previous limited sample studies have been deficient in key aspects such as samples of main-sequence stars, thorough completeness measures in search techniques, and a comprehensive understanding of systematic effects and contamination. \cite{lacourse_2016_59494} reported an absence of Boyajian star analogs among 315,000 stars in the Kepler/K2 fields, despite the large presence of known systematics that must be carefully removed \citep{2016ApJ...818..109A}. The largest search to date was done by \cite{2019ApJ...880L...7S} who conducted a blind search for Boyajian star analogs using the Northern Sky Variable Survey (NSVS; \citet{2004AJ....127.2436W}) and the All Sky Automated Survey for Supernovae (ASAS-SN; \citet{2017PASP..129j4502K}) looking through 2.3 million stars. From 200 initial candidates, they identified 21 dipper stars across the main-sequence and red giant branch, with varying dimming timescales classified as "rapid" and "slow". The largest challenge in interpreting the \cite{2019ApJ...880L...7S} results lies in determining whether these identified main-sequence dippers are true analogs to the Boyajian star. Several critical factors remain unaddressed, including: searches for infrared excess measurements that would indicate warm circumstellar material, detailed characterization of dip durations and amplitudes comparable to Boyajian's star, rigorous vetting of photometric data for systematic effects, and analysis of any underlying periodic signals that might distinguish these events from the complex aperiodic dimming characteristic of Boyajian's star. Without comprehensive assessments, the classification of the candidates reported as true Boyajian's star analogs remains uncertain. Other variables such as R Coronae Borealis Stars \citep{2021ApJ...910..132K}, massive stars with mass-loss events \citep{2021Natur.594..365M}, and planetary debris around white dwarfs (e.g., \cite{2021ApJ...917...41V}), can also exhibit stochastic dimming events. Thus, careful attention is needed to systematically select regions in the color-magnitude diagram to conduct careful searches and understand the underlying selected stellar population. While uncertainties remain about NSVS sample completeness, contamination, and other possible camera systematic artifacts, their work demonstrated that such dimming phenomena are rare, with a similar occurrence rate of $\sim$1 in a million \citep{2019ApJ...880L...7S} for main-sequence solar-like stars. As investigated on a case-by-case basis, searching the origins of such dimming phenomena around main-sequence stars is still a big challenge, and requires extensive multi-wavelength observational campaigns to place meaningful interpretations on what drives their variability. The increasing discovery rate of stochastic dimming events seen among main-sequence stars calls to action for a systematic search to possibly shed more insight into the underlying mechanisms. 

In this study, we present a systematic search for main-sequence dipper stars using multiwavelength photometric surveys. Our analysis is based on a carefully selected sample of high-cadence light curves from the Zwicky Transient Facility, combined with infrared archival data to characterize the long-term variability of these systems. We investigate the underlying physical mechanisms driving the observed dimming events and assess their potential classifications. This paper is organized as follows. In Section \ref{sec:data}, we describe the data acquisition process, including the selection criteria and crossmatching with auxiliary surveys. Section \ref{sec:methods} outlines our dipper search pipeline, detailing the methods used for event identification, classification, and the mitigation of systematics. Next, in Section \ref{sec:Properties}, we analyze the properties of our identified dipper candidates, such as timescales, depths, color properties, and other auxiliary infrared/H$\alpha$ data. Section \ref{sec:discussion} provides a discussion of the potential classification scenarios, including eclipsing binaries, young stellar objects, cataclysmic variables, and occultations by circumstellar material, including disks. Finally, in Section \ref{sec:conclusion}, we summarize our findings and discuss future directions of our study.

\section{Photometric Data} \label{sec:data}

\subsection{ZTF Ubercalibrated}
The Zwicky Transient Facility (ZTF) is a wide-field time-domain survey utilizing the 48-inch Schmidt telescope at Palomar Observatory. It covers the northern sky in g, r, and i bands with a 47 square-degree field of view, achieving median 5$\sigma$ limiting magnitudes of g$\sim$20.8, r$\sim$20.6, and i$\sim$19.9 mag in 30-second exposures \citep{DekanyNew, GrahamZTF, ztf_paper1, MasciZTF}. For our dipper detection pipeline, we use the ZTF Ubercalibration (Zubercal) DR16 photometry that spans observations from March 2018 to March 2023. Zubercal is a new set of photometry based on a detailed recalibration of the ZTF PSF photometry that achieves 1$\%$ or better photometric accuracy. The current ZTF PSF calibration is based on the comparison of each quadrant image with a set of calibration stars selected from Pan-STARRS1 (PS1; \citet{ps1}). However, certain limitations arise, such as the crowding, available PS1 calibration stars, and Galactic reddening at the line of sight, which will overall reduce the accuracy of the estimated zero points and thus the photometric accuracy. The Zubercal pipeline also corrects for varying spatial, chromatic, and temporal structures in the photometry that are inherent due to the CCD inhomogeneities \citep{ZTF_Zubercal}. Zubercal works to our advantage for performing such large photometric searches since it achieves significantly higher accuracy photometry compared to the reported nominal ZTF PSF photometry \citep{ZTF_Zubercal}. In addition, Zubercal comes with more photometry from the ZTF CCD edges since it does not require a ZTF reference frame and compensates for areas with little to no reference frames. Due to the high accuracy, removal of temporal systematics, and additional detections, Zubercal is a well-suited data product to help us achieve a systematic search for dipper main-sequence stars. All dipper candidates investigated in this work are compared against the standardized NASA/IPAC Infrared Science Archive\footnote{\href{https://irsa.ipac.caltech.edu}{https://irsa.ipac.caltech.edu}} ZTF photometry and ZTF science images \citep{IRSA_ZTF_LC} to ensure that no other unwanted systematics are present.

\section{Methods} \label{sec:methods}

\subsection{Gaia FGK Dwarf Selection}\label{sec:fgk_sel}
The selection of FGK dwarfs represents optimal targets for detecting main-sequence dipping events due to their relatively stable photometric behavior compared to more active variable stellar populations. Using early Kepler data, \citet{2011AJ....141..108C} demonstrated that G dwarfs exhibit the lower photometric scatter, with a dispersion floor near 0.04 mmag, while F and K dwarfs show somewhat higher variability, typically a few mmag on the timescale of weeks due to rotational modulation, magnetic activity, and stellar pulsation. On longer timescales of years, studies of magnetic activity cycles of FGK dwarfs have found typical variability amplitudes of 10-30 mmag with an average timescale of 6-8 years \citep{2016A&A...595A..12S}. 

In this work, we utilize $\texttt{StarHorse 2022}$ \citep{2022A&A...658A..91A}, a Bayesian isochrone-fitting framework developed to derive robust stellar astrophysical parameters, including distances and extinction values. Using a Bayesian statistical framework, $\texttt{StarHorse 2022}$ combines observational data with prior information about Galactic metallicity, geometry, and interstellar extinction.\@ $\texttt{StarHorse 2022}$ delivers a catalog of 362 million stars from Gaia EDR3 \citep{2021A&A...649A...1G}, cross-matched with photometric surveys like PS1, SkyMapper \citep{2019PASA...36...33O}, Two Micron All Sky Survey (2MASS; \citet{2mass}), and WISE. The derived stellar parameters of $\texttt{StarHorse 2022}$ are validated through comparisons with other similar works such as \citet{BailerJones21}, \citet{2018A&A...616A...8A}, \citet{2020A&A...638A..76Q}, as well as through open cluster data, astroseismic measurements, and spectroscopic observations. Findings suggest systematic errors are smaller than the nominal uncertainties for the majority of objects, making it a well-suited catalog to select most of our FGK main-sequence stars. 

\begin{figure}
    \centering
    \includegraphics[width=1\linewidth]{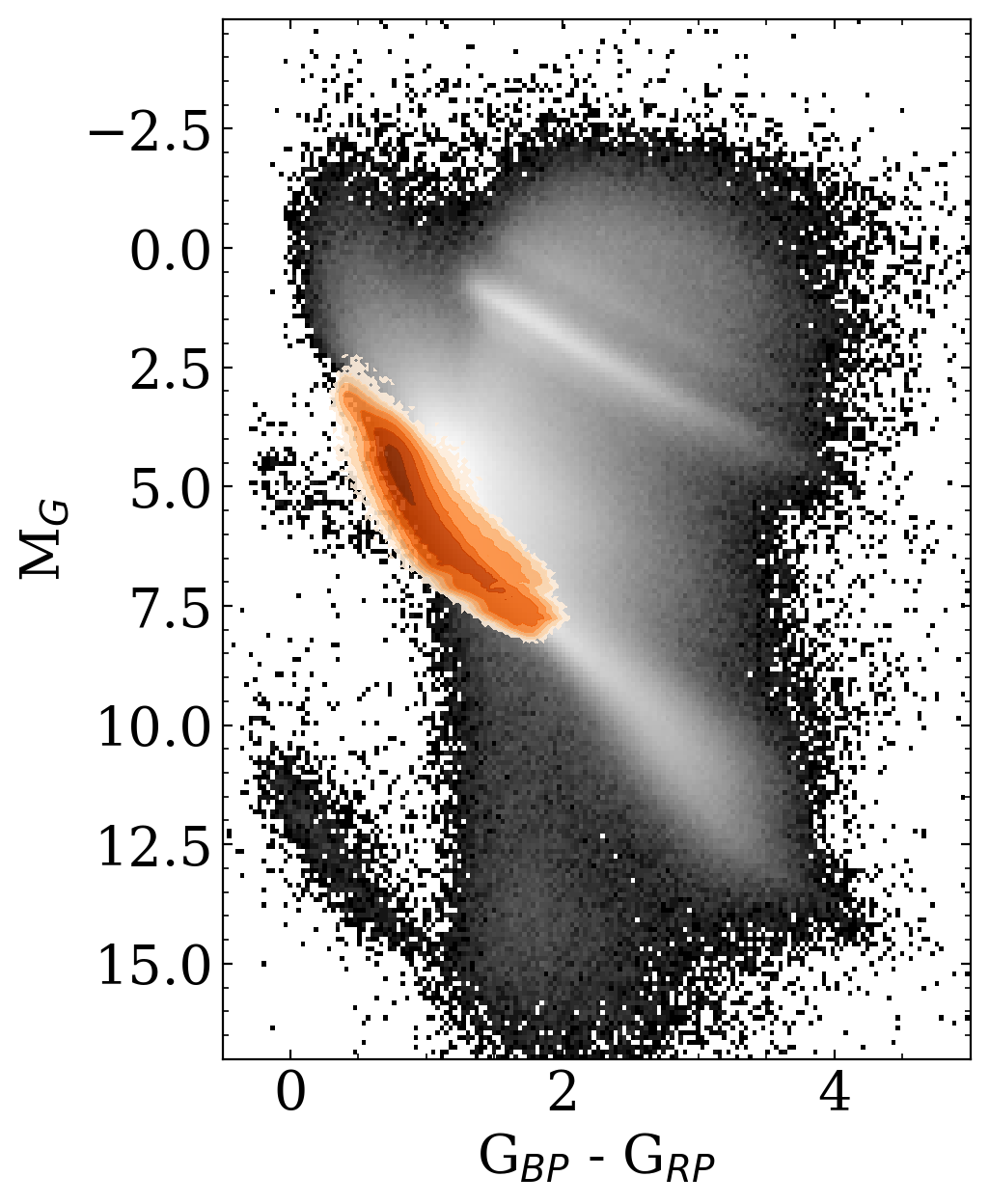}
    \caption{Gaia DR3 color-magnitude diagram (in gray) for the nearby Gaia DR3 sample \citep{gdr3_offc} shaded by the number of stars per bin. Overplotted in the orange contours is the color-magnitude diagram selection of the FGK main-sequence stars using $\texttt{StarHorse 2022}$ \citep{2022A&A...658A..91A} used for this study.}
    \label{fig:gaia_starhorse_cmd}
\end{figure}

First, we selected all $\texttt{StarHorse 2022}$ sources with declination above -30 degrees to match the footprint of the ZTF survey. Next, we select $\texttt{StarHorse 2022}$ sources with log($g$) between 4 and 5, and T$_{\text{eff}}$ between 4000 Kelvin and 7220 Kelvin \citep{2013ApJS..208....9P}. In addition, we impose an astrometric criterion to be $\texttt{fidelity}$ $>$0.9 from $\texttt{StarHorse 2022}$ to ensure that our sample excludes sources with unreliable astrometric solutions (see Figure 15 \citet{2022MNRAS.510.2597R}). We apply this cut to significantly reduce the contamination from objects whose positions in the color-magnitude diagram are compromised due to poor Gaia astrometric measurements and thus increase the reliability of our FGK stellar selection. The query yielded $\sim$106 million stars. Next, we crossmatched these stars to the Pan-STARRS1 survey, assuming a 1-arcsecond nearest neighbor crossmatch that resulted in $\sim$70 million stars. To further ensure we have selected a strict sample of stars on the stellar locus, we have selected sources that only fell within 2$\sigma$ of the PS1 color-color stellar locus in the $(g-r)$ and $(r-i)$, as any outliers in the PS1 stellar locus were likely due to unwanted systematics. Our selected stellar locus was visually compared to the stellar locus of \citet{2007AJ....134.2398C}, and found strong agreement, although we raise the possibility of a few contaminating M-dwarfs at the redder (g-r)$\sim$1.2 colors \citep{2014MNRAS.440.3430D}. In \nameref{ape:appendix1B}, we present our stellar locus selection compared to the work of \citet{2007AJ....134.2398C}. The final resulting catalog contains 63 million candidate FGK main-sequence stars that are to be used for this analysis. In Figure \ref{fig:gaia_starhorse_cmd}, we present the color-magnitude diagram of our final selected sample of FGK main-sequence stars.

\subsection{Finding Dips} \label{sec:finding_dips}

Here we describe how we conduct our systematic search for main-sequence dipper stars. First, we match our selected  FGK dwarfs object catalog (see Section \ref{sec:fgk_sel}) to the Zubercal photometric catalog to access the photometry for each source. We reject all sources with fewer than 50 detections in each ZTF $gr$ photometric filter. Crossmatching is conducted using the Large Survey Database\footnote{\href{https://lsdb.io/}{lsdb.io}} (\texttt{LSDB}) \citep{2025arXiv250102103C} framework that enables rapid crossmatch on large astronomical catalogs leveraging Hierarchical Adaptive Tiling Scheme (HATS) data format for survey structure\footnote{\href{https://hipscat.readthedocs.io/en/stable/}{hats.readthedocs.io}}. To scale our computations of all Zubercal light curves, we utilize the $\texttt{Nested Pandas}$$\footnote{\href{https://nested-pandas.readthedocs.io/}{https://nested-pandas.readthedocs.io/}}$ package for large-scale time-series computations. We used $\texttt{Dask}$\footnote{\href{https://docs.dask.org/}{https://docs.dask.org/}} to implement a CPU-bound parallelization of the computing tasks using 25 cores.

For each star, we compute the biweight magnitude deviation of their ZTF $gr$ light curves defined as: 
\begin{equation}
    \delta_{i} = \frac{m_{i} - R}{\sqrt{\sigma_{m_i}^2 + S^2}},
\end{equation}
where R and S are the robust biweight location and standard deviation of each light curve, respectively, using the $\texttt{astropy}$ implementation. Since we are looking for large amplitude deviations in our time series data, the selection of the biweight is intentional, as it is more robust against outliers than traditional measures such as the mean and standard deviation, which can be strongly influenced by outliers. 

Next, using the magnitude deviation of the ZTF r-band light curves, we search for peaks above a threshold $\delta >3$
using the SciPy \texttt{find\_peaks} utility, which finds local minima by comparing neighboring values of a given array \citep{2020NatMe..17..261V}. To avoid the detection of redundant peaks in our running deviation metric, we enforce a minimum separation of 50 ZTF detections, which corresponds to a time interval of approximately 100-150 days, assuming a typical ZTF cadence of 2-3 days. This criterion is designed to reduce false positives from unnecessary detections of structured dips with multiple minima within the $\sim$150 day interval. In a few cases, our criterion may result in sources being flagged to contain simply one dip as compared to more; however, it should not affect the identification of at least one dip. The peak-finding algorithm returns the times at each identified peak. To eliminate spurious dips from systematic effects, we enforce that a corresponding dip must also be present in the g-band data within two days of the r-band magnitude deviation peak. To identify significant dips, we compare the deviation at the median time of the r-band dip with the corresponding g-band detections within the defined dipper window, and calculate the z-score statistic between the two deviation estimates measured for the g and r light curves. A dip is considered significant if the z-score exceeds $1 \sigma$. While we acknowledge that this threshold is relatively low, through trial and error, we found that it best minimizes bias against events with moderate color dependence. Finally, each identified peak in the ZTF-r band is fitted using a Gaussian function with mean ($\mu$) and standard deviation ($\sigma$). 

\subsection{Dipper Scoring Metric}\label{sec:score_metric}

Various dipper score metrics have been carried out throughout the literature to identify dips in time-series data. For example, \citet{2014AJ....147...82C} introduced the flux asymmetry score that measures the asymmetry in flux fractions of a given light curve. Asymmetry flux-related metrics have been used successfully to identify YSO variability, some of which have stochastic variability, such as the signal this study wishes to address. Other more sophisticated machine learning techniques have been utilized before to find anomalous ZTF light curves using a series of machine learning outlier techniques \citep{2021MNRAS.502.5147M} that have been successful in the past in identifying anomalous light curves. Such workflows have previously benefited from the computationally cheap time-series metrics capable of identifying interesting light curves. We found that previous methods do not take into account the frequency of dips, duration, and the number of available data points at the signal of interest. Such parameters are key in deriving a robust metric that will allow the identification of good candidate main-sequence dipper stars in sparse and noisy light curves such as ZTF. In this work, our aimed scope is not necessarily to discover ``anomalous'' light curves but instead to build a metric that quantifies the number of dips that occur in each light curve as a function of amplitude, duration, and available data (i.e., data quality) over the full light curve history. We introduce a novel light curve dipper score metric as: 
\begin{equation}
    \mathcal{S} = \frac{1}{ln(N+1)^N} \sum_{n=i}^{N} \bigg{(}\frac{\delta_{i}}{2} \times \Phi_{i} \times N_{det, i} \times \frac{1}{\chi_i^2} \bigg{)},
\end{equation}
where we define the number of dips (N), the deviation at the peak of the dip ($\delta_{i}$) as defined in eqn (1), the full-width half-maximum (FWHM) at each peak ($\Phi_{i}$), number of r-band detections within the dip window (N$_{det}$), and the measured chi-square statistic of our Gaussian fit ($\chi^2$). We define the dip edges (i.e., window) by locating the nearest points before and after the dip center where the light curve magnitude first returns to within half a standard deviation below the mean magnitude. We note this metric is for the entire ZTF light curve. Our heuristic scoring metric is designed to capture the full dipper history of a given light curve and rewards light curves with an overall small number of dips conditioned on the quality of the data (i.e., number of data available, and duration of the event). Our light curve scoring metric has a complicated selection function, in nature, due to the noisy ZTF light curves. It is possible that a deep-short dipper can be scored similarly with a shallow-long dip, depending on the amplitude, duration, and number of available points in the duration of the dip. Since we also impose a chi-square penalty on the Gaussian fit, systems that are asymmetric will generally have slightly smaller scores compared to those with symmetric light curves. Throughout the analysis, we discovered events with very small $\chi^2$ fits are likely bogus outliers with only one detection that has good agreement with a Gaussian; we penalized these events by implementing criteria that reduce or negate their score contribution when they exhibit narrow profiles (FWHM $<$ 1.5 days), minimal deviation from the mean magnitude, or suspiciously perfect model fits. To test our light curve score metric in Section ~\ref{sec:injection-recovery}, we perform an injection-recovery analysis to understand the limitations and biases of our scoring system.

\subsection{Addressing ZTF Systematics} \label{sec:systematics_ztf}

\begin{figure}
    \centering    \includegraphics[width=1\linewidth]{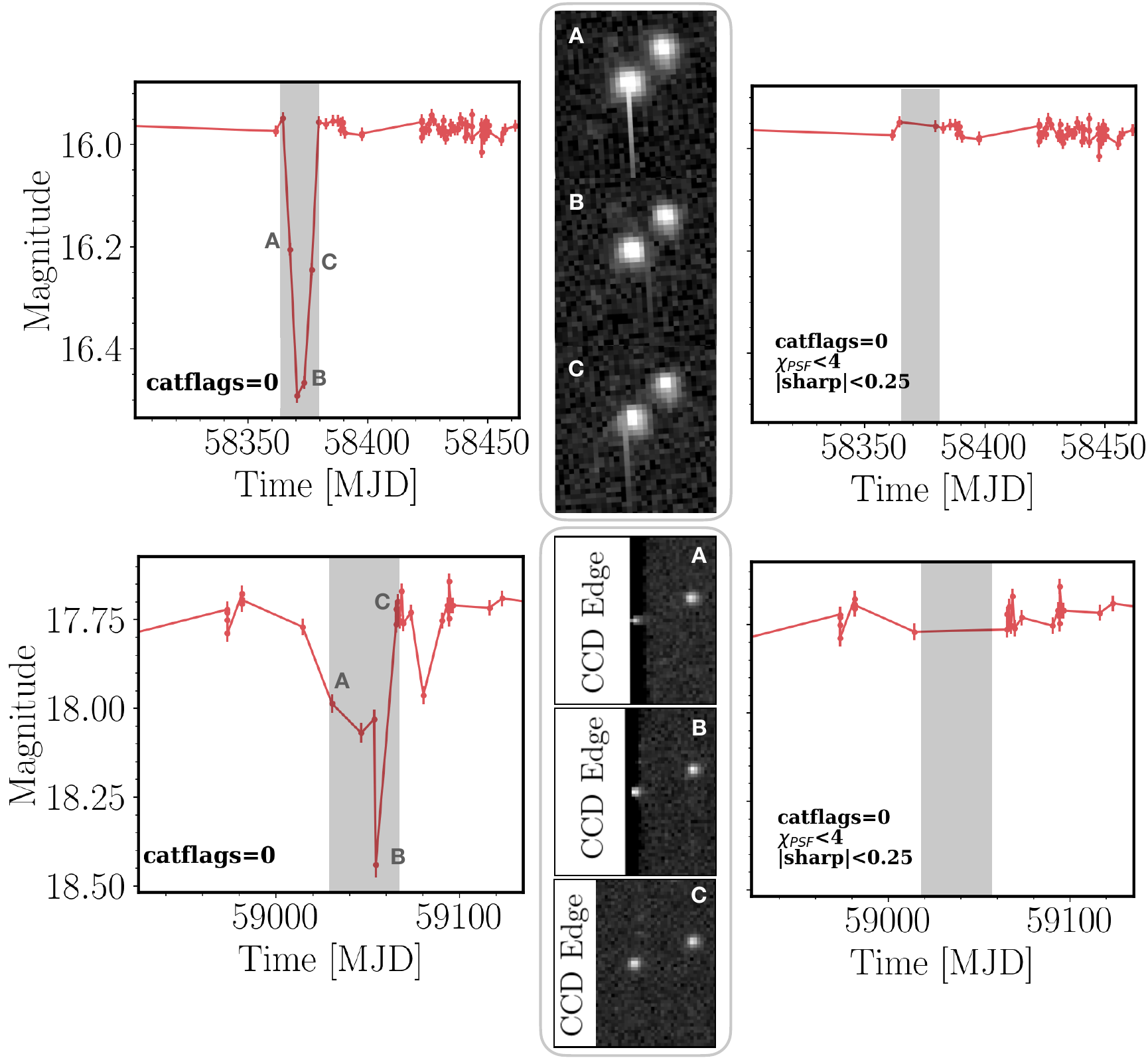}
    \caption{Examples of the two most common types of spurious dips in the ZTF r-band photometry: (top) OID 744211200016789 and (bottom) OID 736206200012619. The first column highlights the prominent dip-like directions with letters A-C, using only the catflags=0 condition. The central column shows the corresponding ZTF science image of each dip detection, where image artifacts are visible. The right column shows the light curve using constraint detections based on catalog-level PSF photometry, which is only available through IRSA.}
    \label{fig:example_fake_dip}
\end{figure}

ZTF photometry exhibits notable and known systematic effects, particularly near the CCD edges, where vignetting and other edge-related issues degrade the photometric accuracy across the \textit{gri} filters, for example, see \citet{2021MNRAS.502.5147M}. Additionally, columns with known defects or objects near saturated sources may cause spurious variability and can result in the misidentification of dipper-like candidate stars. Current bulk data releases of ZTF light curves, including Zubercal, do not provide PSF catalog-level information for individual detections nor any spatial constraints relative to the CCD edges. The absence of key parameters in the ZTF bulk release light curves, to assess the quality of the PSF of each source, including its position on the CCD, complicates the application of a straightforward spatial-photometric cut. In the ZTF Science Data System Advisories and Cautionary Notes\footnote{\href{https://irsa.ipac.caltech.edu/data/ZTF/docs/ztf_extended_cautionary_notes.pdf}{ZTF Science Data System (ZSDS) Advisories and Cautionary Notes}}, for instance, the PSF chi, and sharp parameters can be used to remove further spurious photometry, however, they are only available via the time-costly IRSA light curve GUI service query and not the bulk data release. In Figure \ref{fig:example_fake_dip}, we display the ZTF-r band light curve of two commonly identified systematic effects via the light curves and ZTF Science images before and after we apply PSF-level queries from IRSA. We note that similar effects take place in the g-band filter with similar durations. Based on their light curves alone, such spurious detections that cause ``dips'' at first inspection can look convincing; however, stricter PSF requirements and inspection of the images provide more important context to reject such spurious events.  

\begin{figure}
    \centering
    \includegraphics[width=1\linewidth]{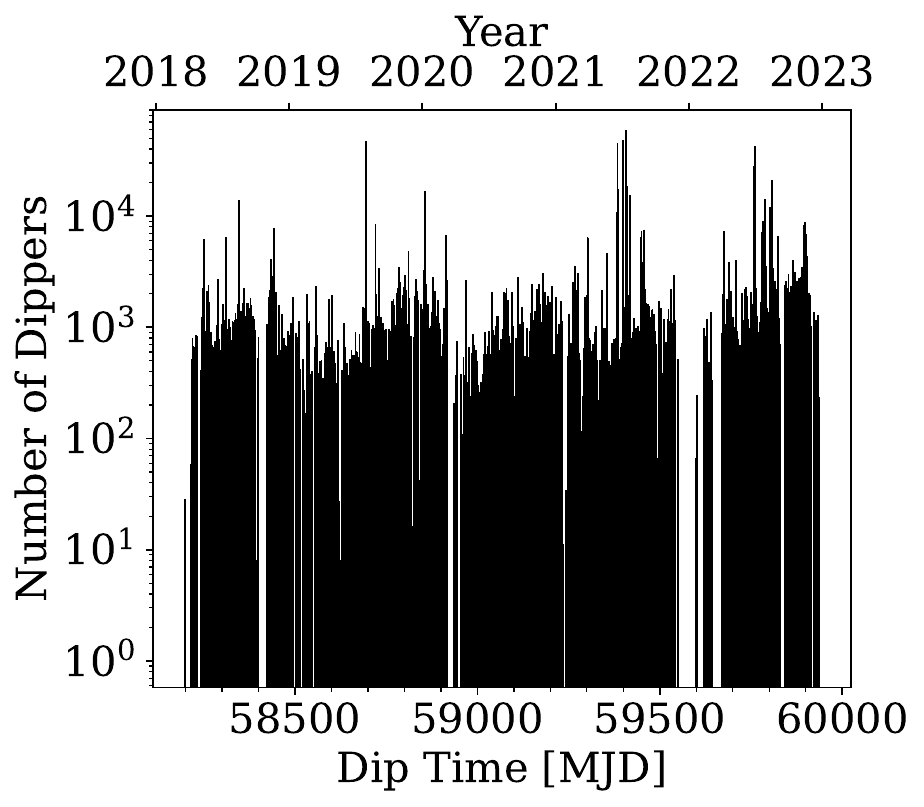}
    \caption{Modified Julian Date time of the highest magnitude deviations for each light curve, binned by a 1-day cadence. The systematics presented are temporary and spatially correlated.}
    \label{fig:dip-time-freq}
\end{figure}

\begin{figure}
    \centering
    \includegraphics[width=1\linewidth]{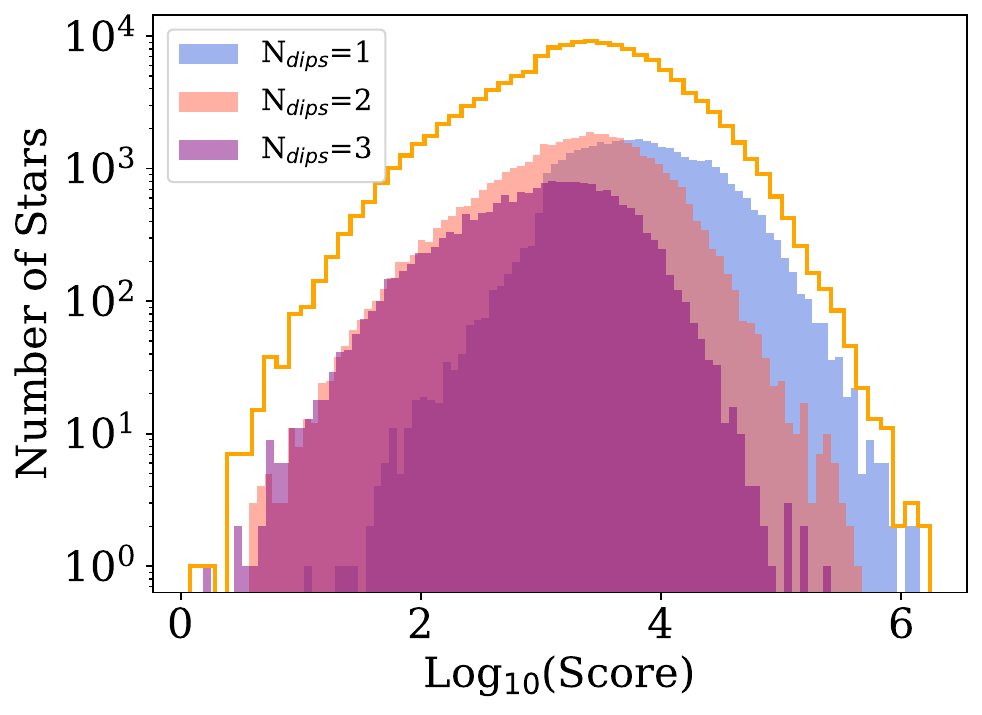}
    \caption{Histogram of the light curve dipper scores after removing bulk systematic errors from the sample. The orange line represents the total score distribution with minimal quality cuts, requiring each dip to have at least one detection at its minimum. The colored histograms, distinguished by the number of significant dips, apply more stringent criteria such that each dip has at least three detections during the dimming phase.}
    \label{fig:lc_score_distribution}
\end{figure}

We investigated the distribution of dip times with the highest amplitude dip that occurred in each r-band light curve shown in Figure \ref{fig:dip-time-freq}. Figure \ref{fig:dip-time-freq} reveals, there are several nights throughout the survey where large overdensities (on the order of 10$^4$ stars) of dips occur within the same night, and have a seasonal pattern. We suspect that this issue specifically arises during the warmest ambient temperature months at the Palomar Observatory, or can generally be attributed to bad weather conditions that cause spurious dips. In the final scoring distribution seen in Figure \ref{fig:lc_score_distribution}, we removed all dips that were within $\pm$0.5 days from the identified overdensity ($>$10$^4$) of stars that we have deemed likely bogus. 

To assess and mitigate the impact of false dipper signals, we analyzed a subsample of 1,000 potential dipper candidates, uniformly sampled from the raw score distribution without filtering for spurious dips. For each candidate, we retrieved IRSA-level photometry and applied more stringent PSF cuts, following the guidelines outlined in the ZTF Science Data System Advisories and Cautionary Notes (see right panel of Figure \ref{fig:example_fake_dip}). After reprocessing the light curves through our scoring pipeline, we found that $\sim$90$\%$ of the sample failed to meet the scoring threshold, as spurious dimming events were effectively removed. The remaining light curves, however, exhibited subtle systematic effects not captured by our PSF cuts, requiring visual inspection to confirm their validity. We conclude that without more specific PSF-level parameters for each detection, we are currently unable to eliminate spurious signals without manual inspection of the science images. While this limitation does not preclude the discovery of our main-sequence dipper stars, we caution that the initial light curve scoring distribution includes a high contamination rate alongside bona fide events. 

\subsection{Injection-Recovery Simulations} \label{sec:injection-recovery}

To evaluate the efficiency of our light curve scoring pipeline, we conducted a basic injection-recovery test using simulated dipper events with the characteristics of the ZTF light curves. We selected a uniform sample of 10,000 stars across the sky footprint and magnitude range from the original 63 million sources that did not meet the initial $\delta > 3$ criterion to select an unbiased control set for our injection-recovery analysis, and avoiding sources that may exhibit significant photometric dips. We modeled each dipper-like event using a skew-normal distribution characterized by parameters for the mean ($\mu$), standard deviation ($\sigma$), skewness ($\alpha$), and offset term (C$_0$). To introduce variability and mimic observational noise, we perturbed the model as follows:
\begin{equation}
   f(\mu, \sigma, \alpha, C_0) + \mathcal{N}(0, \sigma_{mag}) + \mathcal{N}(0, \sigma_{\sigma_{mag}}),
\end{equation}
where $\sigma_{\text{mag}}$ represents the standard deviation of the light curve, and $\sigma_{\sigma_{\text{mag}}}$ is the standard deviation of the magnitude errors. To account for the dependency of magnitude errors on brightness, we estimated a 5th-order polynomial to approximate the average uncertainty as a function of magnitude for all 10,000 of our initial ZTF light curves. The modeled dimming event was then added to the observed data to produce the final simulated light curve. For simplicity, we assumed the dimming event amplitude to be identical in the \textit{gr} bands.

We acknowledge that our simple skew-normal injection model does not physically simulate the underlying mechanisms driving the optical dimming events. This reflects the broader challenge that, to date, no study has conducted (i) a systematic search or (ii) a comprehensive characterization of light curve dimming profile shapes, which are likely influenced by complex and competing processes. For example, Boyajian's star and ASASSN-21qj have exhibited complex dimming profiles that deviate significantly from symmetric Gaussian shapes, while other dipper stars such as HD 166191 \citep{2022ApJ...927..135S} show Gaussian-like symmetric dimming events in their optical light curves. Other optical dimming events have been characterized by the sum of multiple superimposed Gaussians with an underlying Poissonian component \citep{2016MNRAS.463.2265S}. Given the diversity in optical light curve phenomenology, the true functional form of the dimming profile remains uncertain. Thus, we recognize that the skew-normal distribution represented in our injection-recovery modeling is only an approximation for these events.

\begin{figure}
    \centering
    \includegraphics[width=1\linewidth]{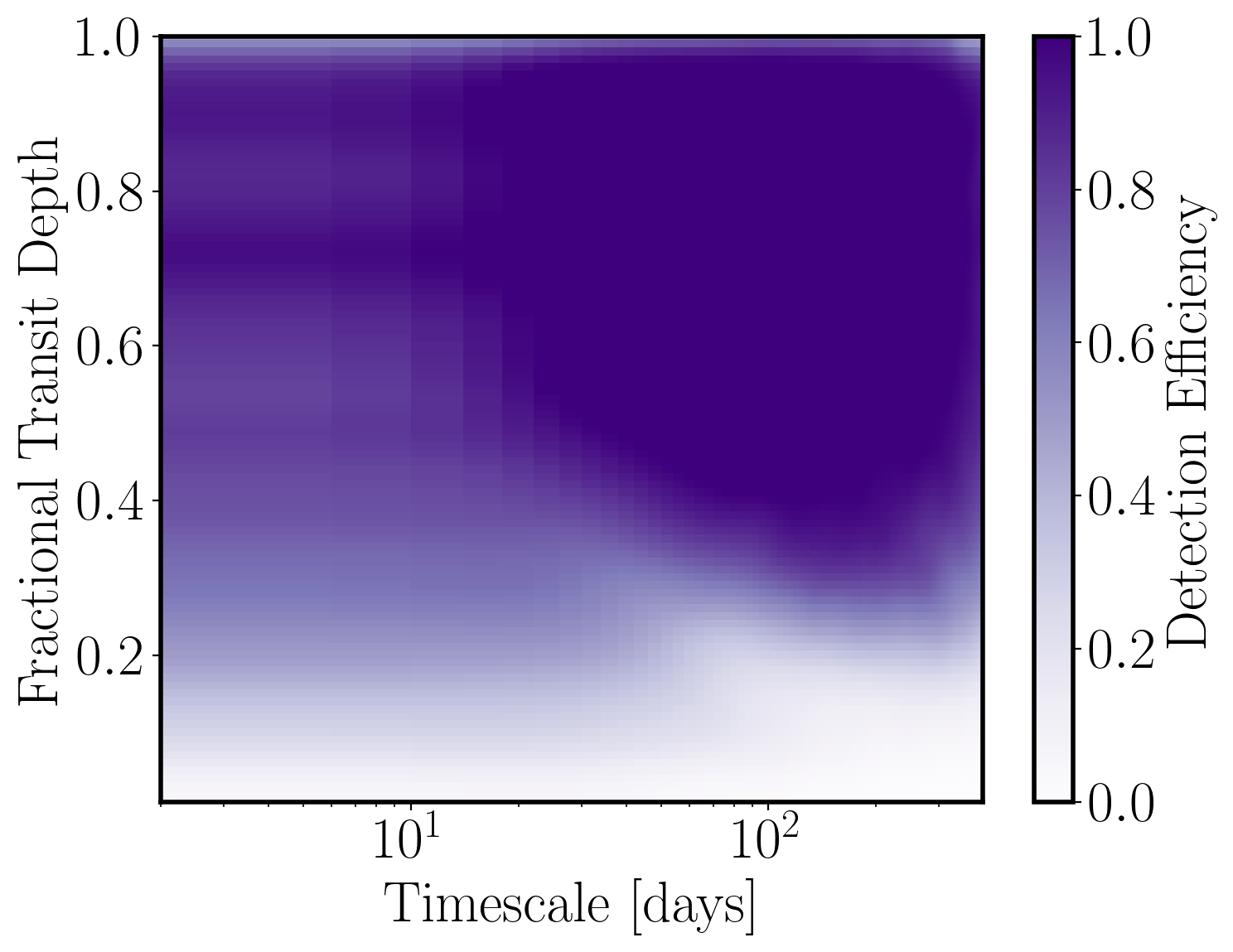}
    \caption{Simulated dipper star amplitude versus duration of the dimming event for our injection-recovery simulations. The color scale represents the median detection efficiency fraction, defined as the ratio of successfully detected events that pass our minimum light curve score criteria to the total number of simulated events per bin. The detection efficiency decreases for shallower and longer-duration dips, reflecting the limitations of our selection criteria and survey cadence.} \label{fig:simulation_efficiency}
\end{figure}

To evaluate our pipeline's performance, we simulated a grid of 100,000 amplitudes and durations to generate synthetic light curves by randomly sampling from the initial 10,000-star dataset. The mean time of the dimming event was randomly sampled from the time axis of the ZTF-r band light curve. For each synthetic light curve, we also drew a skewness parameter value randomly between -0.5 to 0.5 to simulate both symmetric and asymmetric dimming. Each light curve was injected with a single simulated dip. We processed these simulated light curves through our scoring pipeline and assessed the fraction of events that met our scoring criteria. Candidates were considered ``detected'' if they had $\log_{10}$(score)$>$3 and at least three detections during the dimming phase. To match observations, we translate each injected light curve by computing its fractional transit depth and timescale in Figure \ref{fig:simulation_efficiency} and estimate a grid of the detection efficiency. Demonstrated in Figure \ref{fig:simulation_efficiency}, we found that our scoring pipeline is most sensitive  ($\sim$100$\%$ completeness) for events over 50 days with fractional transit depths above 40$\%$. For shorter timescales around 10 days or less, we found that our detection efficiency varies between 20-60$\%$.

Besides the detection efficiency, we also investigated the score distribution of our synthetic dipper light curves and other key summary statistics. For example, from the above example, we found the score distribution to be tightly marginalized between 8$>$log$_{10}$(score)$>$3, which all contained a minimum of three detections in the dimming phase in the ZTF-r band light curve. The marginalized bounds of the light curve score ultimately reflect the cadence and noise properties of the ZTF light curves. In addition, we computed other auxiliary time-series statistics, such as the skewness, Von Neumann Ratio \citep{10.1214/aoms/1177731677, Neu2} that measures the correlated variance and variance of each light curve that can inform our quality cuts based on the real ZTF light curves. 

\subsection{Discovery Process} \label{sec:discovery}

As discussed in the previous section, our first pass at removing ZTF systematics is presented in the light curve log$_{10}$(score) distribution presented in the orange histogram of Figure \ref{fig:lc_score_distribution}. Based on our findings from the injection-recovery testing, we restrict our sample to consider potential candidates as those with log$_{10}$(score)$>$3, must contain at least three detections within the identified dip, and have a skewness value above 0.6. With these first cuts, our sample is cut down to 57,729 candidates to investigate (23,185 one dip, 25,153 two dips, 9,391 three dips). We expect that these candidates remain dominated by false dips due to systematics (Section~\ref{sec:systematics_ztf}). Taking our estimated contamination rate of $\sim$90$\%$ would yield approximately 5,773 stars to investigate as true candidates that deserve more attention. We noticed that in our 57,729 candidate sample, there was still correlated temporal and positional structure across the sky, likely reflected by the untreated systematic effects present in the light curves. To further remove any systematic errors that are temporally and positionally correlated, we computed a spherical sky distance between each pairwise star\footnote{{We define the distance as \footnotesize\[
\bold{D} = \sqrt{r^2 + r'^2 - 2rr'\left(\sin\delta \sin\delta' \cos(\alpha - \alpha') + \cos\delta \cos\delta'\right)}
\]
}, where \textbf{r} and \textbf{r'} are the dip times of each source, multiplied by the speed of light (i.e., $r = c_0 \cdot t_{dip}$), resulting in a sky distance that accounts for the temporal distance between each dip. The terms $\alpha, \alpha'$ and $\delta, \delta'$ refer to the sky coordinates of each pair of sources.} in our sample. We used the closest pairwise distance in this distribution as a threshold, removing sources that were unusually close in this 3D space that we believe are temporally and spatially correlated.

The remaining 21,607 candidates were visually inspected to check for dipper signals and any unwanted systematic errors. Each source was inspected with the SNAD Viewer tool \citep{2023PASP..135b4503M} to investigate each candidate. The SNAD Viewer provides interactive access to ZTF light curves, science images, and complementary data from other surveys. We visually inspected the dimming events in both ZTF \textit{gr} science images found throughout our search and flagged sources that appeared as bona fide candidates. We repeated this process for events with up to three identified dips. Out of the $\sim$20,000 candidates, we have identified 81 promising candidates that we deem as main-sequence dippers. 

We are aware of the shortcomings of this approach in combining both quality cuts and visual inspection, which may yield inconsistencies and biases toward certain variability signals. It is beyond the scope of this study to understand the complex selection function at play since we are dominated by systematic effects, as has been found in other studies \citep{2021MNRAS.502.5147M}. For example, \citet{2021MNRAS.502.5147M} showed a simple rule to potentially identify systematics in the ZTF light curves by noticing that they generally have higher reduced $\chi^2$ and low period amplitudes. Future studies should consider prior treatment of the ZTF light curves to identify common light curve trends per ZTF field via techniques such as Cotrending Basis Vectors \citep{2012PASP..124.1073P}.

\section{Properties} \label{sec:Properties}

For the analysis of our 81 newly identified main-sequence dipper stars, we downloaded the most recent ZTF gr-band\footnote{We exclude the ZTF i-band data since none of our dips contained any i-band detections.} ZTF 22 photometry directly from IRSA. To remove any spurious photometry, we applied the standard photometric quality criteria (i.e. $\texttt{chi}>4$, $\lvert \texttt{sharp} \lvert <0.25$, and $\texttt{catflags}=0$) recommended by the ZSDS as discussed in detail in Section \ref{sec:systematics_ztf}. We provide Table \ref{table:final_candidate_table} that contains the associated coordinates, magnitudes, distance, time of dip, and number of dips in the ZTF light curves.

\subsection{Timescales and Depths} \label{sec:lc-char}

\begin{figure*}
\includegraphics[width=0.8\textwidth]{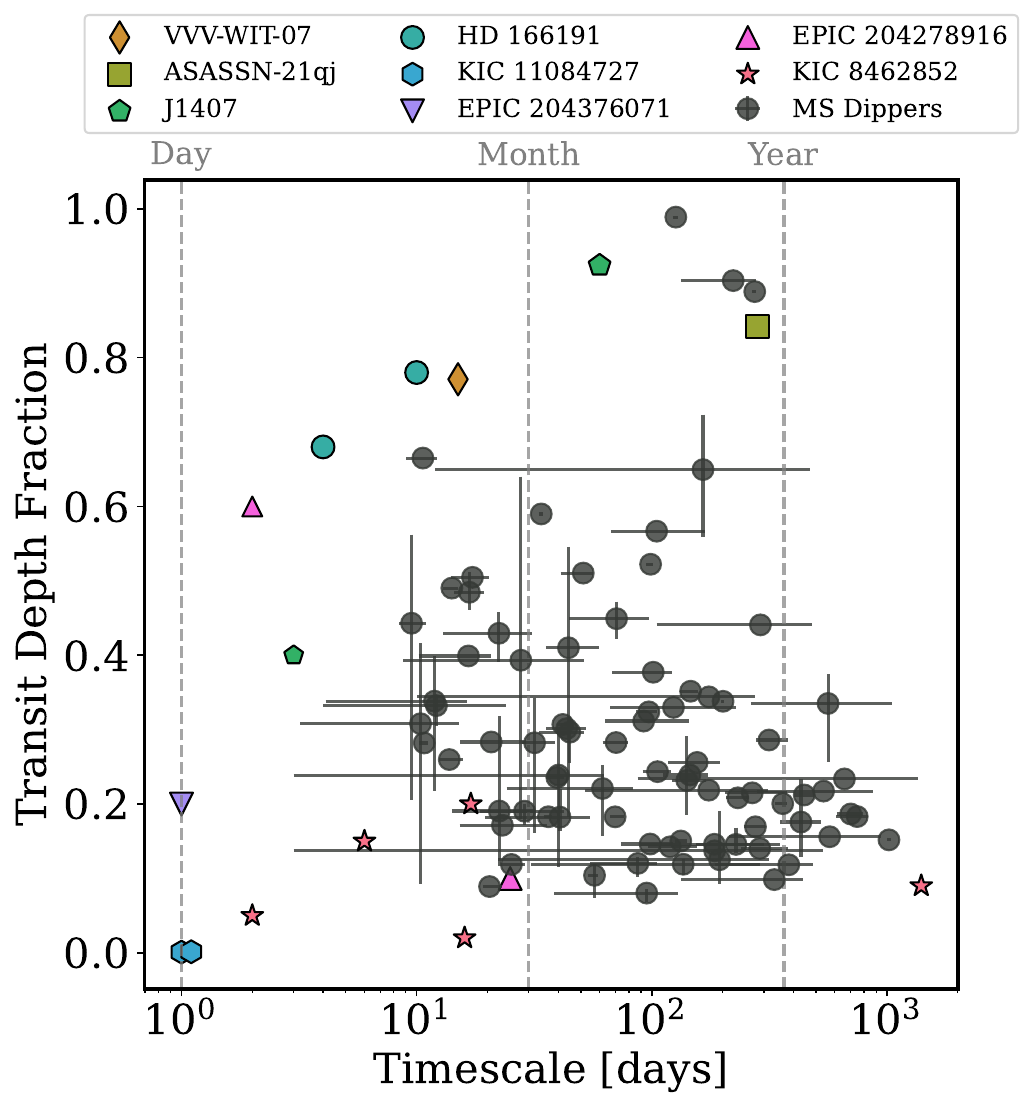}
  \centering 
  \caption{Fractional transit depth versus dimming timescale for main-sequence dipper stars. The fractional transit depth is defined as one minus the minimum normalized flux. Black points represent data from this work, and error bars indicate the range of dimming durations and depths for sources with multiple dimming events. The dashed lines indicate the timescale of days, months, and years. For comparison, data from analogous dipper-like phenomena reported in the literature are included: VVV-WIT-07 \citep{2019MNRAS.482.5000S}, ASASSN-21qj \citep{2023Natur.622..251K, 2023ApJ...954..140M}, J1407 \citep{2012AJ....143...72M}, HD 166191 \citep{2022ApJ...927..135S}, KIC 11084727 \citep{2018MNRAS.474.1453R}, EPIC 204376071 \citep{2019MNRAS.485.2681R}, EPIC 204278916 \citep{2016MNRAS.463.2265S}, and KIC 8462852 \citep{2016MNRAS.457.3988B}.}
  \label{fig:ms_dipper_timescale_depth}
\end{figure*}

The photometric behavior of our main-sequence dipper sample exhibits diverse light curve morphologies that provide important clues into the underlying occulting mechanisms. Through systematic analysis of the ZTF light curves, we identify several distinct categories of dimming events characterized by their temporal evolution, amplitude, and shape symmetry. Complex and irregular stellar dimming events have been observed across various systems, including VVV-WIT-07 \citep{2019MNRAS.482.5000S}, ASASSN-21qj \citep{2023Natur.622..251K, 2023ApJ...954..140M}, J1407 \citep{2012AJ....143...72M}, EPIC 204278916 \citep{2016MNRAS.463.2265S}, and KIC 8462852 \citep{2016MNRAS.457.3988B}. These events have revealed diverse occulting phenomena, including giant ring systems \citep{2015ApJ...800..126K}, tidally disrupted circumstellar disks \citep{2013AJ....146..112R}, dust clumps in YSO accretion disks \citep{2003A&A...409..169B, Bodman_Dippers_YSO}, and planetesimal interactions \citep{2010A&A...524A...8G}. Our search yielded a total of 81 dipper candidates, which we categorized based on the frequency of observed dipping events: 62 candidates (76.5$\%$) exhibited a single dip, 11 candidates (13.6$\%$) showed two distinct dips, and 8 candidates (9.9$\%$) displayed three dipping events during the photometric baseline conducted for this study. Table \ref{tab:lc_morphology} summarizes the number of light curves classified as symmetric, asymmetric, and erratic, determined through visual inspection of their ZTF light curves. In this work, we present ZTF \textit{gr}-band normalized flux light curves from our newly identified main-sequence dipper stars presented in the \nameref{ape:appendix1A}, Figures \ref{fig:comp_lc_single_1}--\ref{fig:comp_lc_multi_2}. The normalized flux is computed as f=10$^{m-M/-2.5}$ where M is the estimated biweight mean of the baseline quiescent light curve.

\begin{table}[ht]
    \centering
    \begin{tabular}{l c c}
        \hline
        \textbf{Morphology Type} & \textbf{Count}  & \textbf{Fraction} \\
        \hline
        Symmetric      & 20 & 0.25 \\
        Asymmetric     & 35 & 0.43 \\
        Erratic        & 26 & 0.32 \\
        \hline
    \end{tabular}
    \caption{Light curve morphology classification based on visual inspection.}
    \label{tab:lc_morphology}
\end{table}

\begin{figure*}
    \centering
\includegraphics[width=0.9\linewidth]{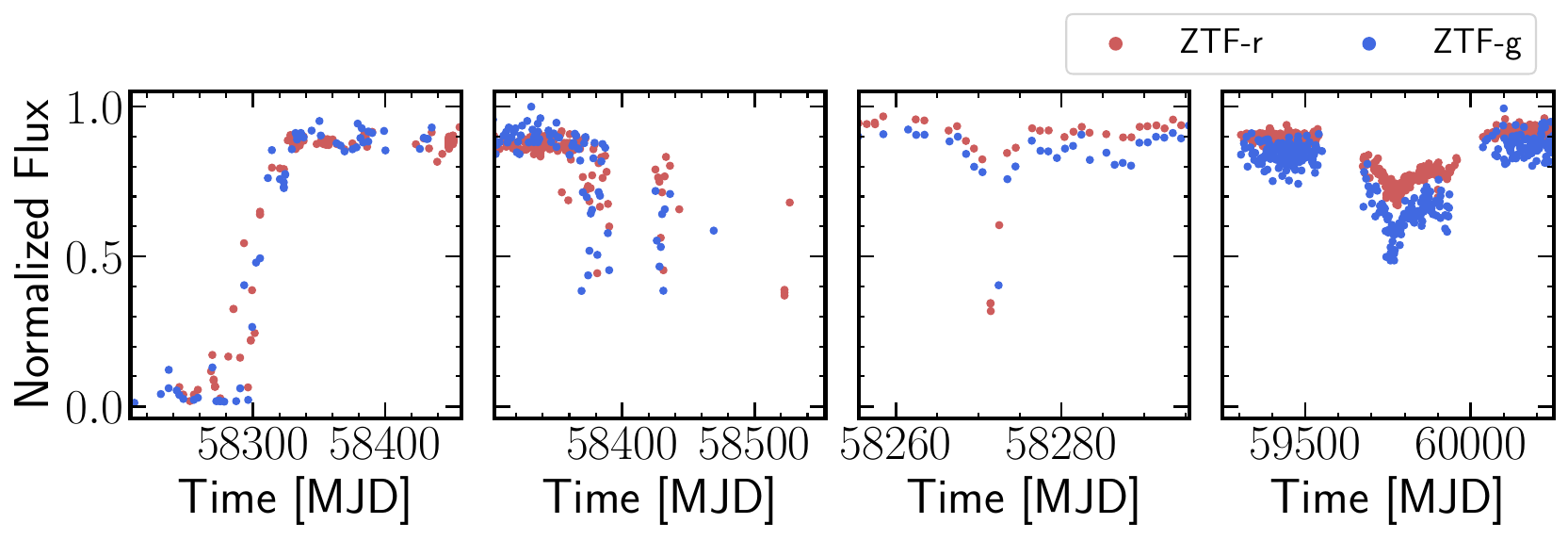}
    \caption{Example trimmed light curves of four identified main-sequence dipper candidates, focused on the identified dips. Each panel represents a different dipper star, showing distinct dimming events with varying depths, durations, and structures. The dips exhibit a range of morphologies and timescales.}
    \label{fig:lc_shape_demo}
\end{figure*}

Our analysis reveals two distinct populations: systems exhibiting single isolated high-amplitude dimming events and those showing multiple dimming episodes. Figure \ref{fig:ms_dipper_timescale_depth} presents the normalized fractional depth versus event timescale for our sample, with fractional depth defined\footnote{Normalized fractional depth is defined relative to the baseline flux level normalized to 1.} at the baseline quiescence of the star. Event durations were determined through visual inspection, marking the start and end time on the epochs of the onset of the ingress and egress of each dip. This deliberate choice was due to ZTF's relatively sparse sampling and the irregular morphology of the dimming events. For context, we include previously reported dipper-like phenomena from the literature. Our light curve timescales span from 3-2000 days, with an average of 100-day duration, with typical fractional depths of $\sim$35$\%$. A small subset ($\sim$6\%) exhibits more extreme characteristics, with fraction depths exceeding 60\% with an average duration of 120 days, suggesting these events are intrinsically rare compared to the low-amplitude-long-duration events. Notably, three of these sources, ZTF18aaypgug \citep{2019TNSTR.246....1D}, ZTF18acebnyy, and ZTF18aboksfn exhibit transit-like features, with fractional transit depths exceeding 90\%. Our findings support the bimodality in event timescales previously suggested by \cite{2019ApJ...880L...7S}, with distinct populations of fast and slow dimming events. The sample displays diverse morphologies, including symmetric, asymmetric, and erratic dimming profiles, as demonstrated more carefully in a selected subset of ZTF light curves shown in Figure \ref{fig:lc_shape_demo}.

\begin{figure}
    \centering
    \includegraphics[width=1.1\linewidth]{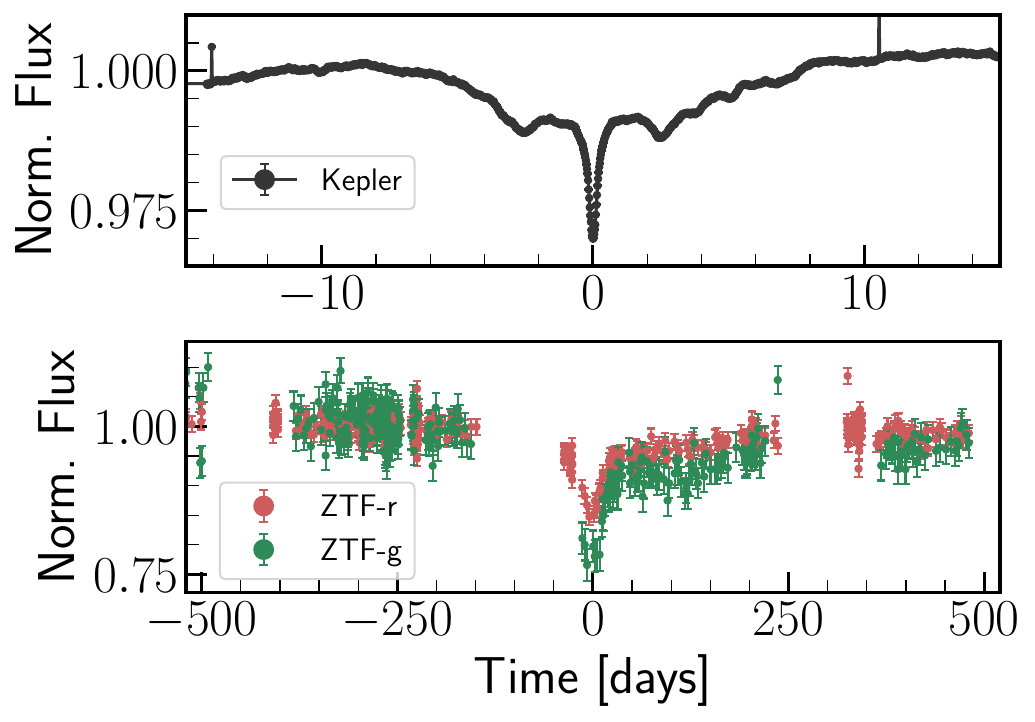}
    \caption{(Top) Normalized Kepler light curve of the Boyajian star (D1500 Dip 9) over a $\pm$10-day interval. (Bottom) Normalized ZTF-$gr$ light curve of ZTF18acqbwiz for comparison, shown over a $\pm$500-day interval. We note the striking resemblance in light curve morphology, though the dip in ZTF18acqbwiz is significantly deeper and longer in duration. Each light curve is shown with its original time baseline to maintain the observed temporal structure.}
    \label{fig:tabby_analog}
\end{figure} 

Our sample exhibits clear selection effects favoring week-to-month timescale events, primarily driven by two factors: our light curve scoring criteria and ZTF's optimal sensitivity window, which is better suited for detecting longer-duration events compared to those approaching the sampling cadence. Nevertheless, the observed distribution of timescales and depths reveals a population of dimming stars spanning broad ranges in both parameters. Notably, extreme dimming events with depths $>$80$\%$ are rare, comprising only 3.7$\%$ of our main-sequence dipper sample. The absence of distinct clustering in the timescale-depth parameter space (Figure \ref{fig:ms_dipper_timescale_depth}) suggests these phenomena occur across a wide temporal range. We found at least 5 candidates that are close to the timescales and depths seen in one of the Boyajian star dips (Dip 8 D1500), and a surprising resemblance in light curve morphology, however, with substantially higher amplitude and duration, for example, shown in Figure \ref{fig:tabby_analog}. The apparent absence of events with depths $<$10$\%$ reflects ZTF's noise floor, particularly at timescales below 3 days, which are constrained by the survey's average two-night cadence.

\subsection{Light Curve Shapes} \label{sec:lc-shape} 

\begin{figure}
    \centering
\includegraphics[width=1.05\linewidth]{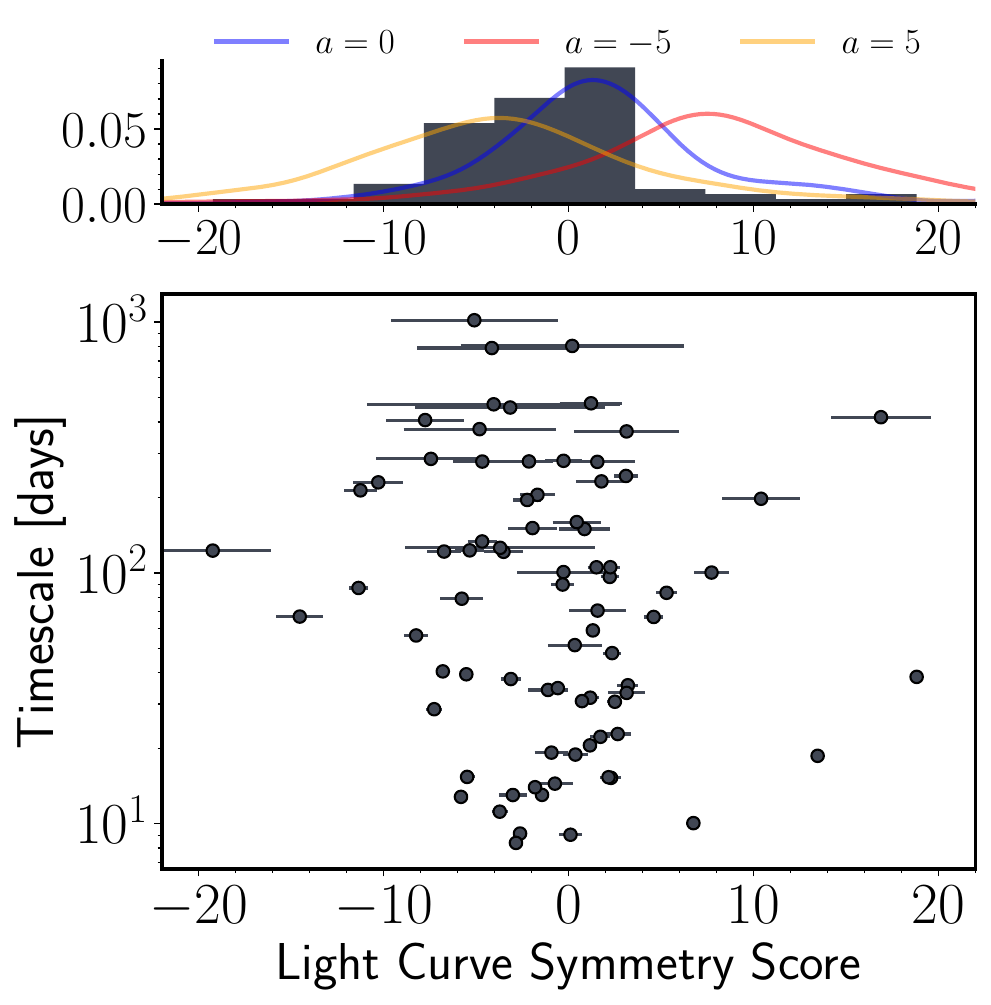}
    \caption{(Bottom) Dipper timescale versus light curve dipper symmetry score. The symmetry score quantifies the temporal structure of each dimming event, where values near zero represent symmetry dimming events. Positive values indicate longer ingress phases, while negative values denote longer egress phases. (Top) Normalized density histogram of light curve dipper symmetry score. Solid colored lines are simulated distributions of dipper light curves with symmetric dimmings (blue; $a=0$), positive skew (orange; $a>0$), and negative skew (red; $a<0$).}
    \label{fig:lc_shape_versions}
\end{figure}

Understanding the light curve shape of the dimming events could help constrain the physical structure and motion of the obscuring material. For instance, asymmetric dimming profiles like those seen example candidate, Figure~\ref{fig:lc_shape_demo}, could be attributed to extended dust tails, warped disks, and circumstellar material on non-circular orbits. We therefore assess the degree of symmetry in each light curve dip to identify potential patterns in our newly discovered main-sequence dipper stars. For each main-sequence dipper light curve dimming event, we measure the light curve dipper symmetry score defined as follows:
\begin{equation}
    \label{eq:integral}
    \mathcal{I}_{i,j} = \sum_{n=i}^j (R_m-f_n) \cdot \frac{t_{n+1} - t_{n-1}}{2},
\end{equation}
where $t_n$ is the time and $f_{n}$ the normalized flux at which observation $n$ was taken, and R$_{m}$ is the robust biweight mean of the normalized flux ZTF-r light curve, which we assume to be the quiescent baseline of the light curve. To measure the symmetry score of a dip, we compare the integrals from a point in the center of the dip at index $c$ to the start and end of the dip window at indices $s$ and $e$, respectively. We calculate the asymmetry score as: 
\begin{equation}
    \label{eq:asymmetry_significance}
    \text{Symmetry Score} = \frac{\mathcal{I}_{s,c} - \mathcal{I}_{c,e}}{\sqrt{\sigma_{\mathcal{I}_{s,c}}^2 + \sigma_{\mathcal{I}_{c,e}}^2}}\ , 
\end{equation}
where we combine the normalized flux of each \textit{gr} light curve, linearly interpolate them, and estimate the asymmetry score as defined from above. In Figure \ref{fig:lc_shape_versions}, we present the light curve dipper symmetry scores against the identified dipper timescales. Events close to zero are generally considered to be symmetric dimming events, while those with positive or negative values will have skewed dimming profiles, such as longer ingress or longer egress, respectively. We find no strong correlation between the light curve symmetry score and dip timescale. However, the overall distribution of symmetry scores (top panel of Figure ~\ref{fig:lc_shape_versions}) shows a modest negative skew of $-0.3$, with a median of $-0.7$, suggesting a small preference for dips with shorter ingress and longer ingress timescales. Since ZTF's irregular cadence can lead to poor sampling, we assess whether this asymmetry could arise purely from sampling effects. Using the simulated light curves in Section ~\ref{sec:injection-recovery}, we simulated 1,000 dipper light curves with symmetric dimming (i.e., skewness parameter $a=0$) with varying amplitude and time of dip, while keeping the duration constant. Following the same light curve symmetry score calculation, we ran a Monte Carlo simulation sampling our synthetic symmetric dimming and sampling them without replacement, and recorded the number of events with negative values. We find that the number of negative symmetry scores in the real data lies $\sim3\sigma$ above the expected mean from our simulations, indicating that the observed negative skewness is moderately unlikely to result from sampling noise. Our tests indicate a statistically significant excess of dips with faster ingress and longer egress, perhaps consistent with the scenario where trailing structures, such as dust clumps with extended trails, can cause leading edge occultations, as we will further explore in Section ~\ref{sec:csm}. 

\subsection{Light Color Evolution Properties} \label{sec:lc-color}
Beyond the amplitude, temporal, and shape variations alone, the temporal color behavior of our main-sequence dippers provides additional constraints on the nature of the obscuring material. Here, we analyze the ZTF-$\textit{gr}$ color evolution of our sample to determine whether significant chromatic changes accompany any dips, and if those changes support reddening scenarios or point to alternative scenarios. 

\begin{figure}
    \centering
    \includegraphics[width=1\linewidth]{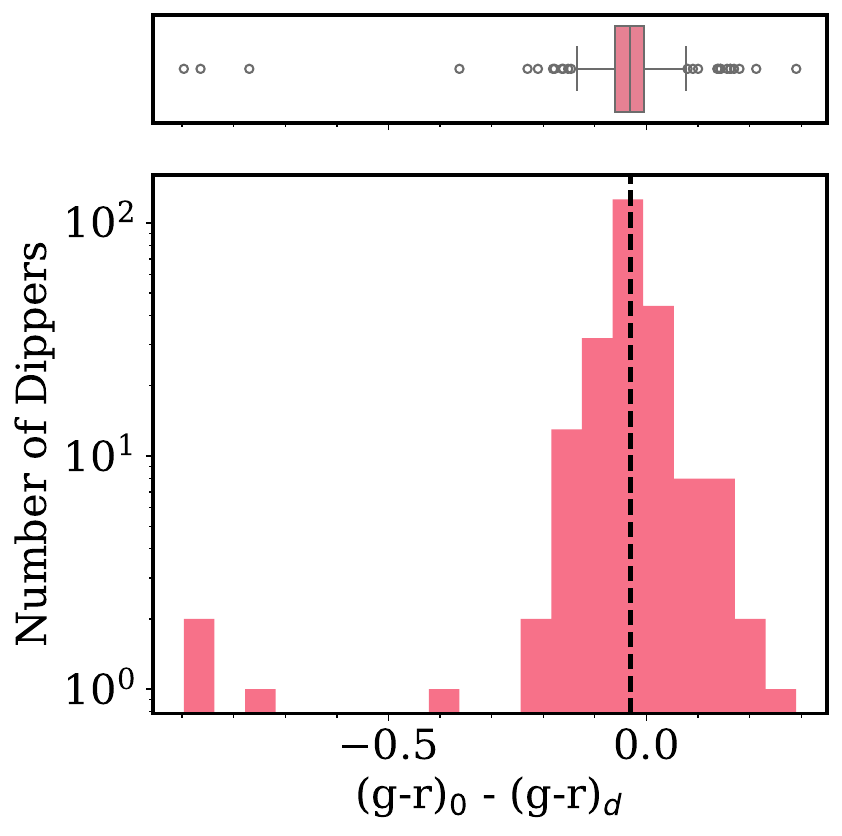}
    \caption{(Bottom) Histogram of the ZTF $(g-r)$ colors for the identified main-sequence dippers. The color excess, defined as $(g-r)_0 - (g-r)_d$, represents the deviation from the mean of the quiescent phase of each light curve, where $(g-r)_d$ is the median color near the centroid of the dimming event. The black dashed line marks the median of the distribution. (Top) A box plot of the color excess is shown at the top of the histogram, illustrating the interquartile range.}
    \label{fig:dipper_gr_colors}
\end{figure}

We analyze the ZTF \textit{gr} color properties of dippers through multiple approaches. First, we examine the color difference between quiescent baseline and dimming events for all the observed dips in their ZTF light curves. Figure \ref{fig:dipper_gr_colors} shows the distribution of ZTF $(g-r)$ color differences between the stellar baseline (measured as the median magnitude outside the dimming events) and during dimming episodes (computed as the mean magnitude during the dip defined in Section \ref{sec:score_metric}). The median color difference is -0.01 mag, with a 2.3 standard deviation that extends to both redder ($<$0) and bluer ($>$0) color changes. We found $77\%$ and $23\%$ of the dips were below and above zero, respectively, suggesting that most dimming exhibited moderate levels of reddening compared to their quiescent baseline. Interestingly, we discovered a significant negative correlation between color change and dip amplitude, with higher amplitude events generally exhibiting redder colors, which likely reflects the extinction effects from the occulting material.

\begin{figure}
    \centering
    \includegraphics[width=0.8\linewidth]{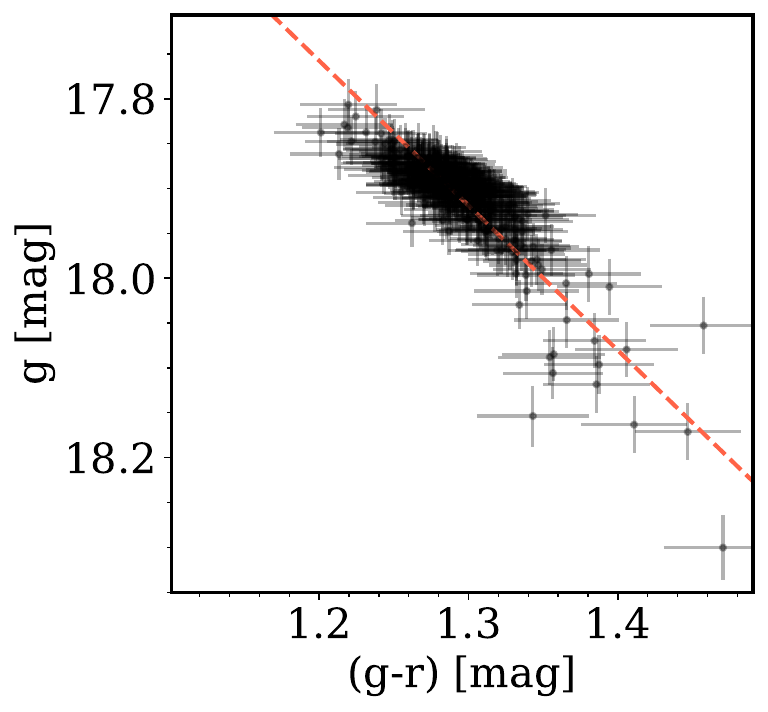}
    \caption{Example of ZTF19aawfjbq ZTF color-magnitude slope. The dashed red line represents the best-fit line.}
    \label{fig:example_gr_fit}
\end{figure} 

\begin{figure}
    \centering
\includegraphics[width=1.05\linewidth]{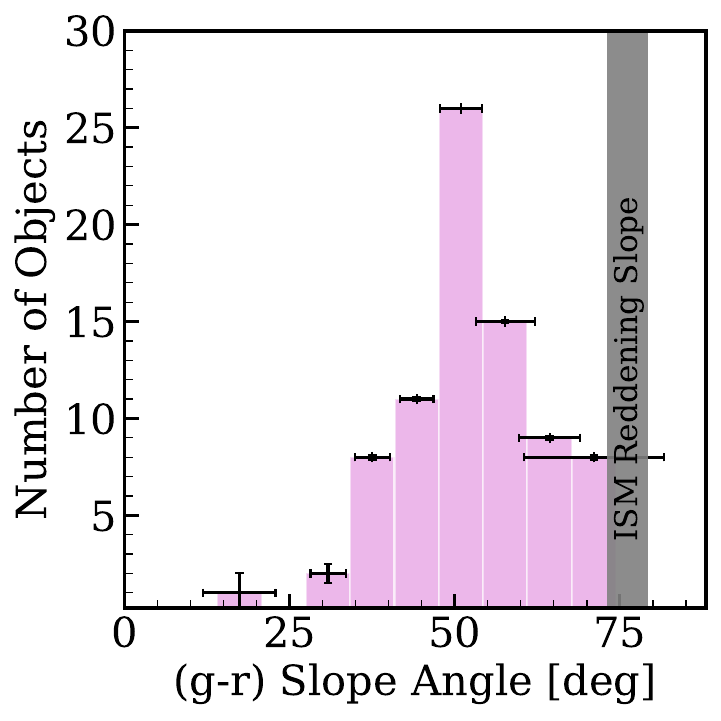}
    \caption{Histogram of the fitted ZTF $(g-r)$ slope angle in degrees, including their standard errors and median slope angle errors. We denote in the gray region, where typical ISM reddening slope values occur between $3<Rv<5$.}
    \label{fig:DipperSlope}
\end{figure}

\begin{figure}
    \centering
    \includegraphics[width=0.9\linewidth]{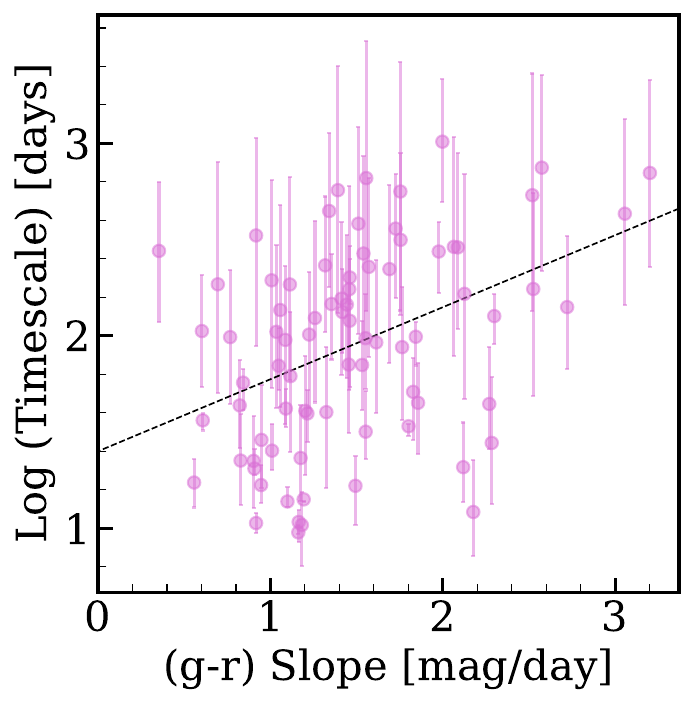}
    \caption{Correlation between event timescale and $(g-r)$ color slope. The obtained data suggest that longer-timescale dimming events generally exhibit steeper $(g-r)$ slopes compared to shorter-timescale events. The black line represents the best-fit line using a least-squares optimization.}
    \label{fig:var-color}
\end{figure}

Next, to investigate the color magnitude diagram (CMD) across our entire observation period, we implemented a modified 3-day running median on the combined light curves. Standard running median methods were found to underestimate the true color behavior. To estimate as closely as possible the true color properties of our light curves, we developed an adaptive binning approach where the maximum magnitude within each 3-day window was selected when the light curve showed excursions exceeding 0.3 mag from the baseline. Following the methods outlined in \citet{2015AJ....150..118P}, for each source, we constructed CMD and quantified their slopes using Orthogonal Distance Regression via \texttt{scipy.odr}, which employs a Levenberg–Marquardt algorithm that accounts for uncertainties in both axes. The slope angles, defined as the inverse tangent of the linear regression coefficients, were compared against typical interstellar medium (ISM) reddening laws \citep{Fitzpatrick99_v1, Fitzpatrick99_v2}. We show a typical fit to the CMD in Figure \ref{fig:example_gr_fit}. To ensure our results were not an artifact of the temporal binning, we repeated the analysis with bin widths of 0.5, 1, and 3 days, obtaining consistent results across all trials. Our analysis reveals that the majority of sources exhibit shallower slopes than those predicted by high-R$V$ ISM extinction curves, where roughly 10\% of our sources are within the margin of typical RSM reddening slopes. The median slope angle in our sample is 54.7 degrees, substantially lower than typical ISM slope angles between 74.1 and 79.2 degrees for R${v}$ values of 3.1 and 5, respectively \citep{ZTF-YSO}. The distribution of CMD slopes is presented in Figure~\ref{fig:DipperSlope}. The observed slope angles suggest that if extinction is responsible for the dimming, the occulting material likely has small R$_v$ values ($<$2), indicating grain sizes smaller than or comparable to the observed wavelength ($\sim0.4\mu$m), though based on CMD slopes, our results seem to indicate that are shallower than other anticipated effects such as hot spots, cool spots, accretion or scattering \citet{2015AJ....150..118P}. We interpret the skewed $(g-r)$ slope angle away from ISM reddening slopes as potential evidence for localized and non-symmetric obscuration events with different geometry or grain properties than typical ISM values.
We detect a moderate but statistically significant correlation (Pearson-R = 0.46, $p<$0.0001) between event timescale and color-magnitude slope, as shown in Figure \ref{fig:var-color}. The global color properties and temporal correlation could indicate an intrinsic relationship between event duration and the dimming properties, though we note that longer-duration events inherently provide a better sampling of color evolution, potentially introducing a measuring bias. 

\vspace{1cm}

\subsection{Demographics} \label{sec:demographics}

The current work contains one of the largest samples of main-sequence dipper stars in a confined color-locus and marginalized stellar parameter phase space. At the moment, it remains unclear if all discovered systems in our sample are indeed driven by the same mechanism of variability (i.e., intrinsic or extrinsic). For simplicity, we will assume that broadly most of our discovered systems belong to the same class of variable stars, for example, systems with one versus multiple dips, or characteristic timescales.

\begin{figure}
    \centering
\includegraphics[width=1\linewidth]{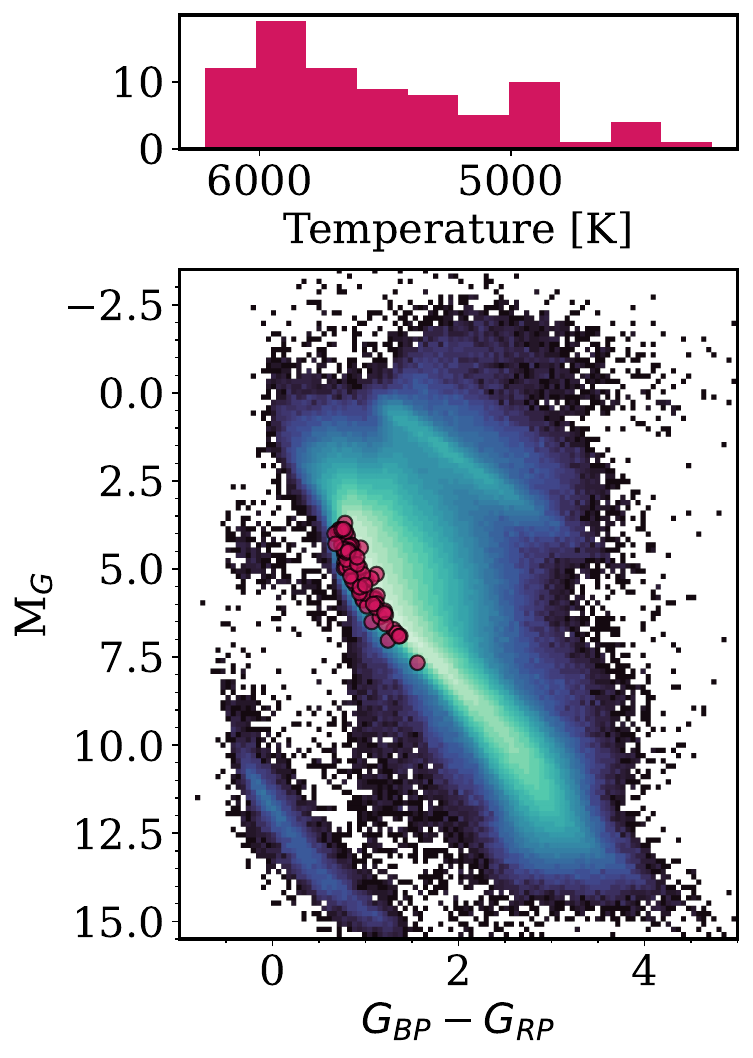}
    \caption{Color-magnitude diagram (blue) for the nearby Gaia DR3 sample \citep{gdr3_offc}, shaded by the number of stars per bin. The positions of our identified main-sequence dipper stars, corrected for extinction using estimates from \citet{2022A&A...658A..91A}, are overplotted. The top panel shows the 50th percentile of the effective temperature posterior distribution for our main-sequence dipper stars obtained by the \texttt{StarHorse 2022} catalog.}
    \label{fig:cmd_diagram_dippers}
\end{figure}

\begin{figure}
    \centering
\includegraphics[width=1\linewidth]{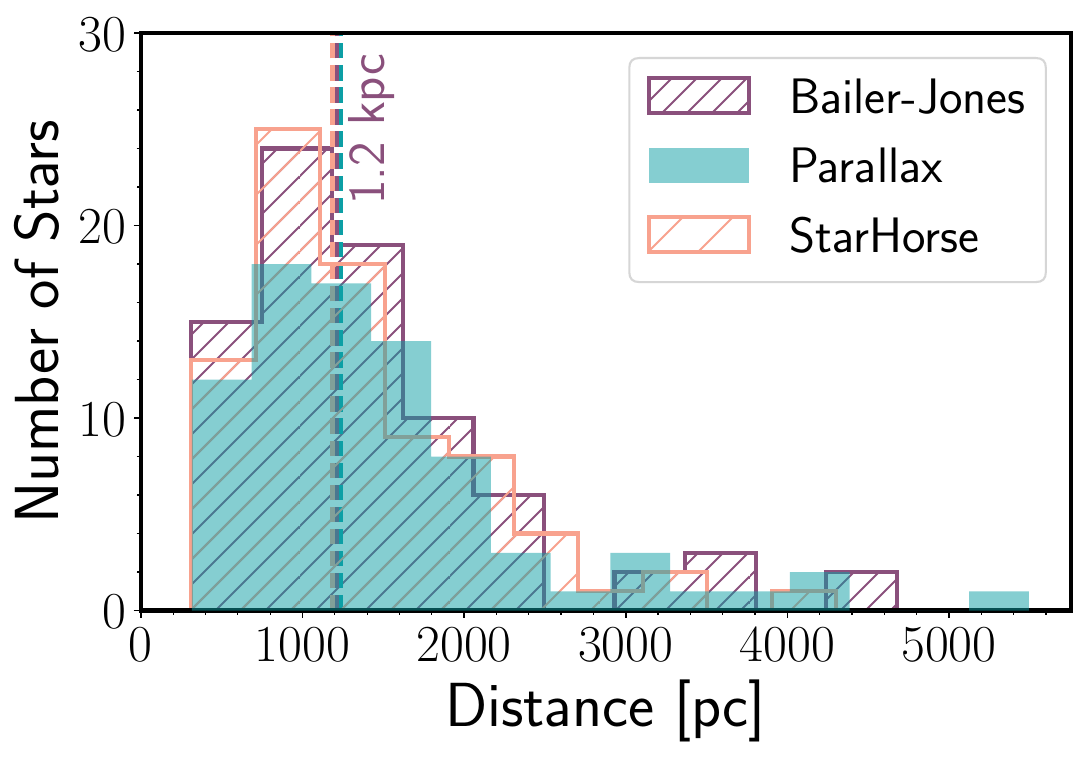}
    \caption{Histogram of the distances to our main-sequence dipper stars, as measured in three separate catalogs: Bailer-Jones \citep{BailerJones21}, parallax \citep{gdr3_offc}, and \texttt{StarHorse 2022}. The dashed lines indicate the median distance for each distribution.}
    \label{fig:dipper_distance}
\end{figure} 

First, we investigate the position of our discovered main-sequence dipper sample on the Gaia color-magnitude diagram. In Figure \ref{fig:cmd_diagram_dippers}, we display a sample of the Gaia color-magnitude diagram (background blue density) and overplot the positions of the main-sequence dippers using distances estimated by $\texttt{StarHorse 2022}$ that have also been corrected for the inferred line-of-sight extinction in the Gaia-G band ($A_{G}$). As expected, our sample lies along a tight locus near where we expect most FGK dwarfs to be found. On the top panel of Figure \ref{fig:cmd_diagram_dippers}, we show the 50$^{th}$ percentile effective temperature of the stellar parameter and notice a slight overdensity of events above $>$5,500 Kelvin. We went back to our initial sample and ran a Kolmogorov-Smirnov (KS) test on the two distributions to investigate if the overdensity in hotter stars was real. We found strong agreement with the two distributions, with p-value $>$0.05, thus not able to reject the null hypothesis and therefore conclude that our bias temperature likely comes from our sample selection criteria. The median distance is 1.2 kpc with an average uncertainty at $G_{mag}<18$ around $30 \%$, confirmed by the Bailer-Jones \citep{BailerJones21}, Parallax \citep{gdr3_offc}, and \texttt{StarHorse 2022} measurements, shown in Figure \ref{fig:dipper_distance}.  

\begin{figure*}
    \centering
    \includegraphics[width=0.9\linewidth]{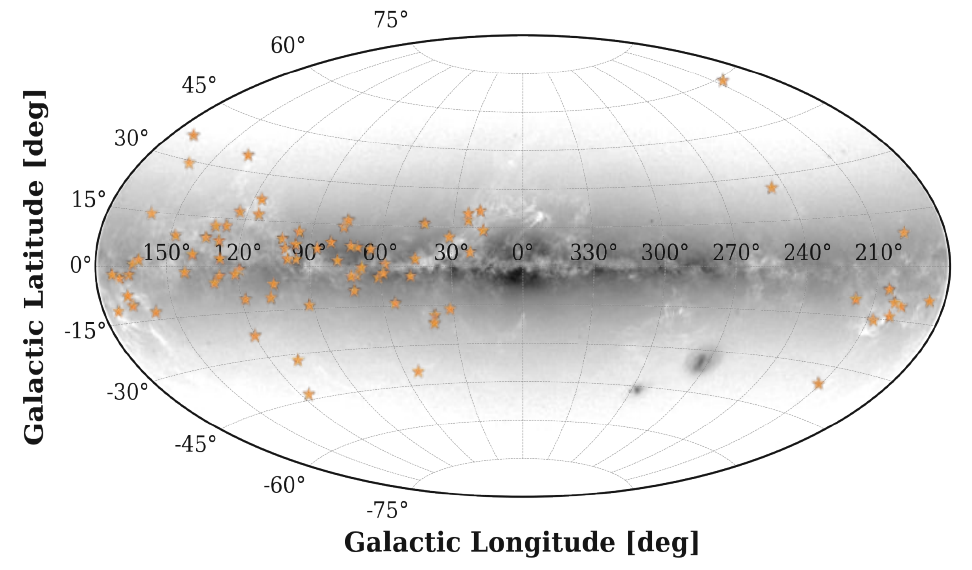}
    \caption{Sky positions of the identified main-sequence dipper candidates (orange points) in Galactic coordinates. The background color map represents the number of Gaia DR3 sources per bin, highlighting the density of stars across the sky.}
    \label{fig:sky_distribution}
\end{figure*}

\begin{figure}
    \centering
    \includegraphics[width=\linewidth]{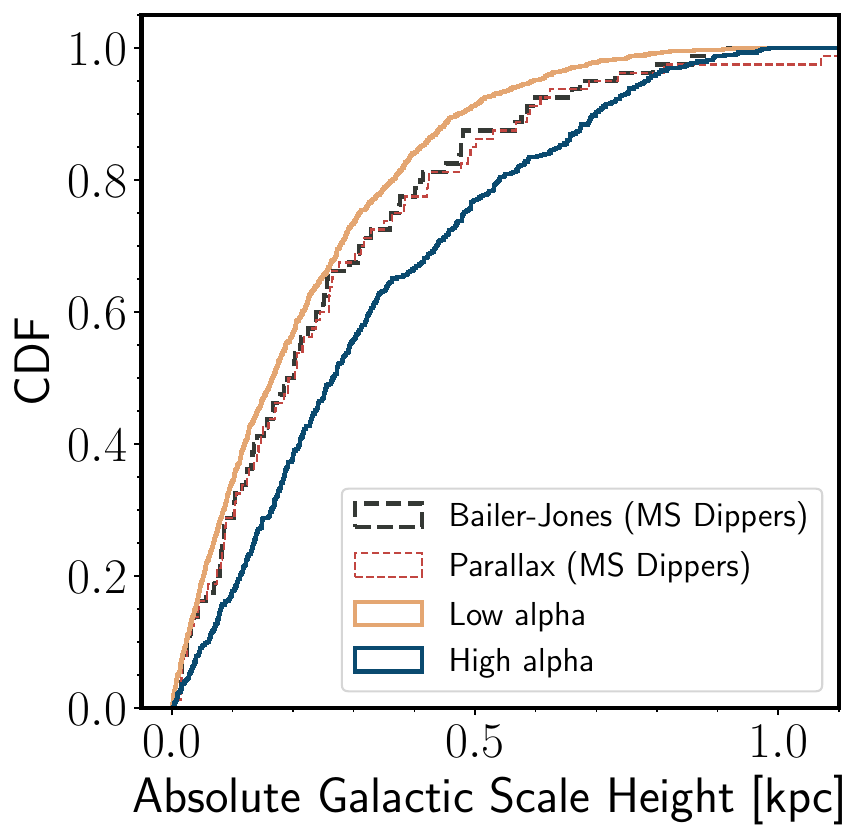}
    \caption{Cumulative distribution function (CDF) of the absolute Galactic scale height for main-sequence dipper candidates, compared with low-alpha and high-alpha disk populations of single stars. The dashed and dotted lines represent the distributions for dipper candidates based on distance estimates from Bailer-Jones \citep{BailerJones21} and the Gaia DR3 parallax measurements, respectively. The solid orange and blue lines correspond to the low-alpha and high-alpha disk populations.}
    \label{fig:scale_height_lot}
\end{figure}

The sky distribution of our main-sequence sample is biased toward the Galactic disk (see Figure \ref{fig:sky_distribution}), where the stellar density is the highest. The median proper motions reported by Gaia DR3 $\mu_{\alpha}$=-0.09 mas yr$^{-1}$, $\mu_{\delta}$=-3.0 mas yr$^{-1}$ with standard deviation 6.7, 5.8 for each vector, respectively. These values are well within the average velocity dispersion for field stars, although without the full 6D (i.e., missing radial velocity measurements), we cannot confidently rule out this scenario.

Given the positions and distance of our main-sequence dippers, we investigated whether any are associated with young stellar associations or star-forming regions. First, we used the BANYAN $\Sigma$ code \citep{2018ApJ...856...23G} that estimates the Bayesian probability of members of young stellar associations within 150 pc of the Sun. We used the reported Gaia DR3 proper motion, parallax, and coordinates of each candidate to assess the BANYAN $\Sigma$ probabilities. For each star in our sample, we found the probability of being a field star is $>99 \%$. We note that the majority of our sources do not contain any radial velocity measurements\footnote{As noted in \citet{2018ApJ...856...23G}, typically the missing quantities (i.e., radial velocity) are marginalized over using Bayesian inference, integrating across the full range of plausible radial velocities at the given position. This approach accounts for the uncertainty introduced by the missing data while still enabling robust membership probability estimates.} Next, we searched using a 1-arcsecond cone search in a homogeneous catalog of 1867 well-characterized stellar clusters using Gaia DR2 astrometry and photometry with cluster distances and ages \citep{refId0} and identified two candidates with well-matched distances. ZTF18abbvemu, which is a member of the ASCC 114 open cluster \citep{Kharchenko} with an estimated age of 7 Myr \citep{2009MNRAS.397....2E}, and ZTF18aboksfn, which is part of the RSG-5 cluster with an estimated age of $\sim$40 Myr \citep{2024ApJ...976..234B, 2016A&A...595A..22R}. We conducted another cone search centered on known star-forming regions and clusters listed in Table 1 of \citet{2022A&A...664A.175P} within $<$1.5 kpc, chosen to approximately match the radius that contains half of the identified members. While 10 of our sources are colocated in proximity to known SFRs, only one appears to be located at a distance consistent with the associated region. This source has been flagged as a potential YSO, ZTF19aayvizm, and may be associated with the Cygnus X star-forming complex, which has an estimated age up to $\sim$10 Myr. At least two other candidates have been identified to belong to open clusters, though with low probability, as discussed in \nameref{ape:appendix1C} Table \ref{tab:viz_table}. Future spectroscopic follow-up programs could fill in missing radial-velocity data to check more concretely for membership probabilities. Stellar age is a particularly interesting proxy for contextualizing the classification observed in stellar variability. For example, young stars can exhibit a spread in photometric morphology \citep{2018AJ....156...71C} across many timescales and amplitudes \citep{2023ASPC..534..355F}. The stellar age of our discovered systems has significant constraints on the progenitor material that must produce a large portion of the occulting material. Stars younger than 20 Myr, for example, may still have evidence of prominent bursts or dust occultation due to accretion and substructure of circumstellar material \citep{TESS_dipper}. We performed another KS test to compare our sample's distribution with the low and high alpha disk populations of the Milky Way within 1 kpc using \texttt{Cogsworth} \citep{Wagg_2025} code. Our results show a strong agreement with the low-alpha disk, with a p-value of 0.59, suggesting that our sample is dominated by younger, low-alpha stars. In contrast, the high-alpha disk comparison yielded a lower p-value of 0.036, indicating a statistically significant deviation from this older stellar population.

\subsection{Periodicity} \label{sec:period_search}

In this section, we discuss the potential implications of discovering periodic sources. To assess the presence of any periodic signal in our discovered systems using a Lomb-Scargle Periodogram (LSP) \citep{1976Ap&SS..39..447L, 1982ApJ...263..835S}. We used the $\texttt{AstroML}$ LSP implementation \citep{astroML}. We run this implementation for the ZTF-r band light curves. To estimate the significance of each LSP iteration, we used a bootstrap simulation of 1,000 simulations by re-drawing from the original light curve with replacement, and computed the 1$\%$ and 5$\%$ significance level of each periodogram. For each computed periodogram, we investigated all peaks that were above the 1$\%$ significance level as potential periodic signals in our data. To exclude possible systematic cadence effects from ZTF light curves, we removed any significant periods that were within a few hours of known alias frequencies and their harmonics close to the sidereal day and cadence of ZTF \citep{2021MNRAS.505.2954C}. For the remaining period candidates that were above the 1$\%$ level and are not near known alias periods, we then took each light curve and phase-folded each at the period to investigate the overall scatter in the light curve. We identified at least one candidate (ZTF19adgqwvh; coordinates (ra, dec)=144.75644, -20.05037, seen in the second column of Figure \ref{fig:comp_lc_multi_2}) that appears to be strongly periodic close to an orbital period of 0.992 days, and we suspect it to be an eclipsing binary close to the ZTF average cadence. For a few candidates, we identified a $\sim$0.1 mag periodic modulation at approximately 29.6 days that could well be within the anticipated nominal stellar rotation of main-sequence stars (e.g., \citet{2013A&A...557L..10N}), though we caution that this period is also very close to the lunar phase period. The majority of our candidates appeared to have no obvious periodic signals that coincide with the observed dips. For additional validation, we conducted a conesearch around a ZTF periodic catalog by \citet{ZTF_periodic}. We found 14 crossmatches from their catalog and visually inspected the phase-folded ZTF light curves at the reported periods by \citet{ZTF_periodic}. In our investigation, we identified one compelling source (ZTFJ213811.15+535115.4), which appears to have a low-duty cycle $\sim$0.1 mag variability at 2.57 day period and has been flagged as a BY Draconis-like variable. The remaining sources appeared to have no coherent structure in their phase-folded light curves. We also inspected the Renormalized Unit Weight Error (RUWE) from Gaia DR3 to infer the likelihood of a companion leading to astrometric excess noise. Generally, systems with an RUWE score of $\sim$1 are consistent with single stars, while thresholds above 1.4 can indicate the presence of a companion \citep{2020AJ....159...19Z, 2020MNRAS.496.1922B}. Out of the 81 candidates, we found 5 stars with RUWE above 1.4 (ZTF18absdnwk, ZTF18actxeox, ZTF18actxeox, Gaia DR3 380715251262643200, Gaia DR3  1084506523872363008) that could potentially host companions. Although \citet{2022RNAAS...6...18F} has suggested that large RUWE scores can also be associated with young stars with circumstellar disks, these 5 systems do not seem to have any particular infrared excess in the 2MASS and WISE bandpasses. However, further high-resolution imaging or radial velocity observations are still required to determine the presence of a potential companion and its properties. 

\subsection{Infrared Properties} \label{sec:ir-properties}

\begin{figure}
    \centering
    \includegraphics[width=1.1\linewidth]{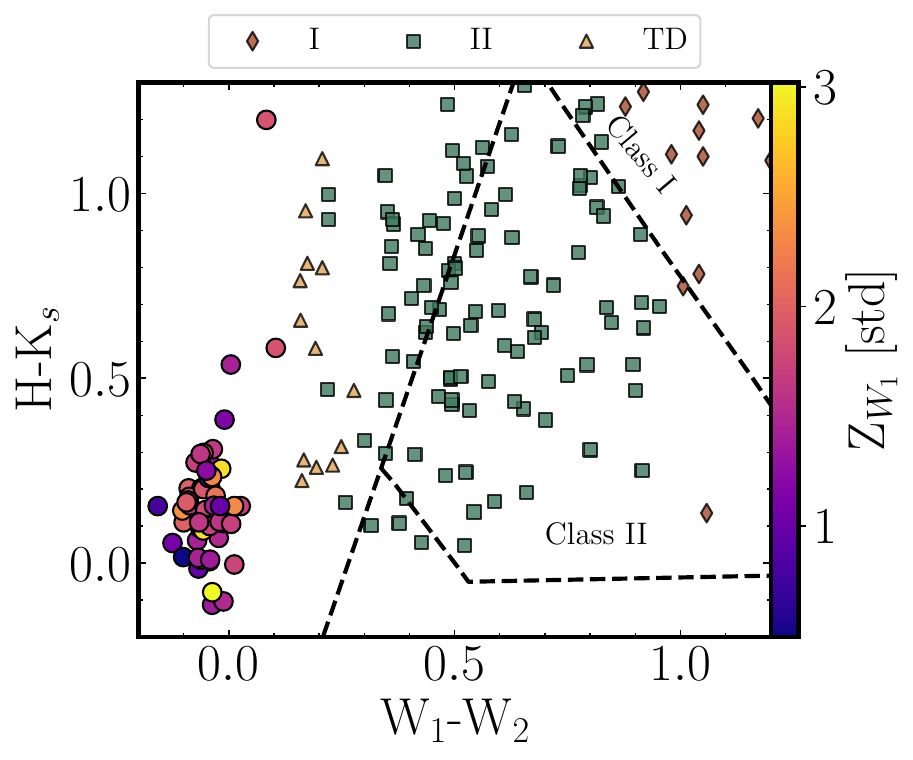}
    \caption{2MASS and WISE color-color diagram of our identified main-sequence dippers, color-coded by the WISE W$_{1}$ variability Z-score indicated in the color bar. The dashed lines indicate the typical boundaries for Class I and Class II young stellar objects, following \citep{2014ApJ...791..131K}. For comparison, we include the 2MASS/WISE colors obtained by \citet{2022ApJ...939...46P}.}
    \label{fig:2mass-wise-var}
\end{figure}

To characterize the infrared properties of our main-sequence dipper stars, we analyzed archival data from 2MASS and WISE. We utilized the unWISE catalog \citep{unWISE} for the WISE data, which is particularly valuable for our candidates located in crowded Galactic fields. The unWISE catalog combines approximately 25 million single-frame WISE images into a deep coadd, enabling more accurate modeling of heavily blended sources in the WISE 3-5$\mu m$ W$_{1}$ and W$_{2}$ bandpasses. To measure the infrared magnitudes with unWISE, we utilized the time-domain data from the unTimely data \citep{2023AJ....165...36M}. Using the ZTF coordinates, we crossmatched to the 2MASS and unTimely catalogs, assuming a 1-arcsecond radius. We found 69 such sources, making up 85$\%$ of our initial sample. To measure the W$_{1}$ and W$_{2}$ magnitudes we took all source detections from unTimely and enforced that the fraction of flux in this object’s PSF that comes from this object is above 50$\%$, the quality factor above 90$\%$, and target distance is less than 3-arcseconds from the initial query coordinates, and take the median of all epochs W$_{1}$ and W$_{2}$ measurements. In Figure \ref{fig:2mass-wise-var}, we show the color-color diagram of our identified stars, including the Z-score variability of the W$_{1}$ unTimely light curves of each star. For comparison, we included the 2MASS-WISE color-color positions of known Class I and Class II, and transition disk young stellar objects discovered in the Sh2-112 Filamentary Cloud Complex \citet{2022ApJ...939...46P}. We marked the boundary separations for Class I, II, and traditional disks following the separation boundary described by \citep{2014ApJ...791..131K}. In Figure \ref{fig:2mass-wise-var}, we found that all 69 sources are well within the main-sequence stellar locus, though we interpret the small W$_{1}$-W$_{2}$ spread for the stellar locus to be due to the effects of metallicity \citep{2019RNAAS...3...54D}. Our search resulted in no obvious sources with bright infrared magnitudes, or any large amounts of time-domain infrared W$_{1}$ variability less than $<3 \sigma$. Within the 2MASS-WISE CMD, it appears that all 69 sources are within the color-color regime of either the main-sequence locus or transitional disks. In addition, we visually inspected the unTimely W$_{1}$ and W$_{2}$ light curves. We did not find any peculiar sources that exhibited infrared brightening or dimming within the identified times of our main-sequence dipper stars within the average 187-day unTimely light curve cadence. It remains unclear if the remaining 12 main-sequence dipper sources have infrared excess or large amounts of variability, and we encourage follow-up. 

\vspace{1cm}

\subsection{Auxiliary Surveys} \label{sec:aux-survey}

\begin{figure}
    \centering
    \includegraphics[width=1\linewidth]{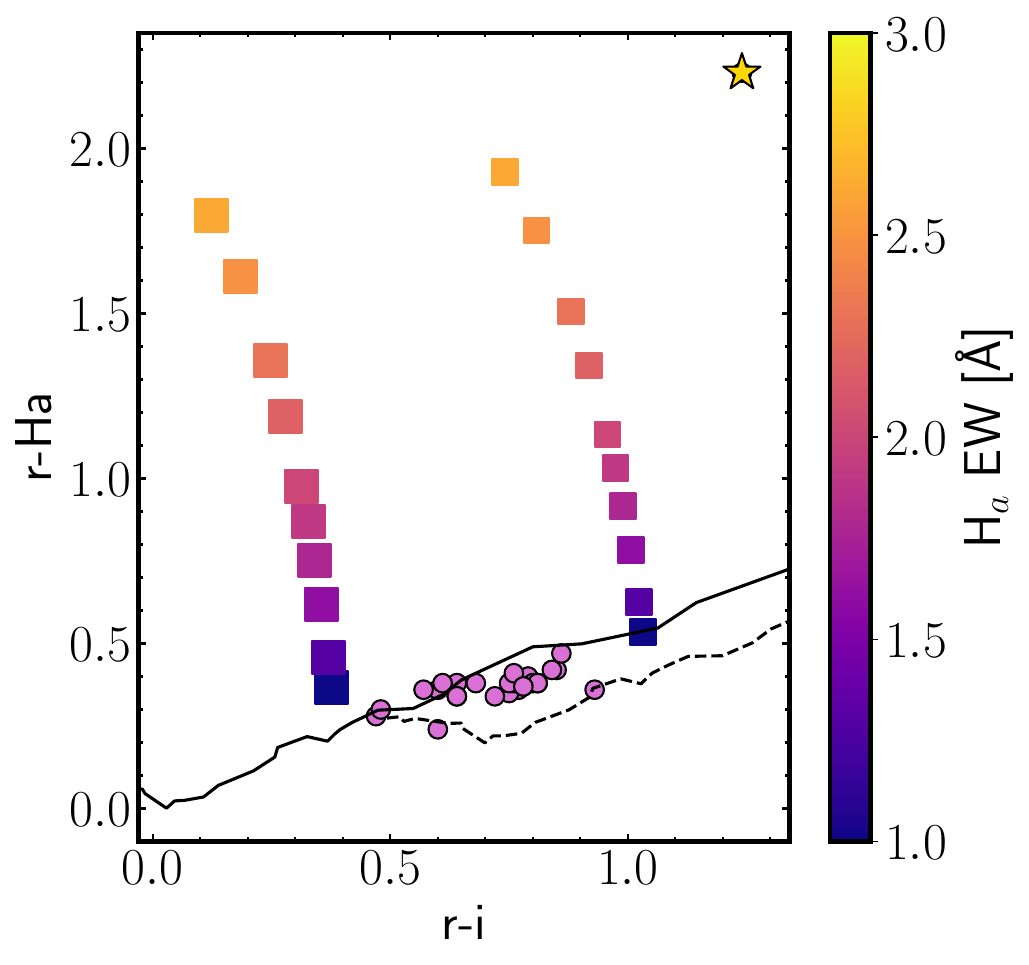}
    \caption{IPHAS color-color diagram showing (r–i) versus (r–H$_\alpha$) for our identified dipper candidates. The pink dots represent the identified main-sequence (MS) dippers, including a bona fide H$\alpha$ emitter (star symbol) in the top right corner. The black solid and dashed lines indicate the main-sequence stellar locus for extinction values of E(B–V) = 0 and E(B–V) = 1, respectively. The square-colored markers represent the predicted color evolution of a G2V star, color-coded by the logarithm of the H$\alpha$ equivalent width (in Å), with separate tracks corresponding to E(B–V) = 0 and E(B–V) = 1. Out of our 81 identified dipper candidates, 33 are found in the IPHAS DR2 catalog.}
    \label{fig:IPHAS_color_color}
\end{figure} 

The identification of H$_{\alpha}$ excess emission can be a telltale sign of the type of variable star that can be driving such anomalous dips. Since most of our sources are in the Galactic plane, we searched to find our sources in the INT/WFC Photometric H$_{\alpha}$ Survey of the Northern Galactic Plane (IPHAS) Data-Release 2 catalog \citep{IPHASDR2}. We crossmatched our stars to the IPHAS DR2 survey with a 1-arcsecond cone search and found 33 sources. In Figure \ref{fig:IPHAS_color_color} we demonstrate the IPHAS color-color diagram of our identified sources with the IPHAS $(r-H_{\alpha})$ and $(r-i)$ photometry. The majority of our stars lie on a locus that is consistent with the predictions of the SED stellar locus of main-sequence stars seen in the solid and dashed line obtained by \citet{IPHASDR1}. Using the relationship between the color-color locus, we show in Figure \ref{fig:IPHAS_color_color} the anticipated equivalent width (EW). Based on the reported IPHAS photometry, only one source was found to have moderate levels of H$_{\alpha}$ emission. As we mentioned, we found one of our systems to have a high excess in the IPHAS $(r-H_{\alpha})$ color, which is interpreted to have the presence of H$_{\alpha}$ emission. We have identified the candidate ZTF18aaypgug (also reported as Gaia19aow \citealp{2019TNSTR.246....1D}). Upon closer inspection of the ZTF-$gr$ light curve of ZTF18aaypgug near the start of 2019, the identified source exhibited a $>$1.5 mag dimming event that lasted for several months. After the 2019 dimming event, the light curve has a significant scatter outside of the dimming event of 0.1 mag in each band. We inspected the ZTF science images and ruled out contaminating neighboring sources (the closest star is $>$5 arcseconds away). We ran several iterations of the LS periodogram on the data by sampling various timelines of the ZTF-r light curve and were unable to find any significant periods between a few hours to 150 days that minimized the initial intrinsic scatter of the phase-folded light curve. By comparing the color-color location of ZTF18aaypgug, we note that these colors are within the range of typical S-type Symbiotic binaries (Sb). S-type Sb systems are interacting long-period binaries composed of a compact hot white dwarf accreting wind from a cool Red Giant Branch (RGB) star companion \citep{2008A&A...480..409C}. The majority of these systems are known for large orbital periods between 200 to 6000 days \citep{2003ASPC..303.....C} and can exhibit complex light curve behavior that involves ellipsoidal variability connected with orbital motion (S-type), radial pulsations, and semi-regular variation of the cool component \citep{2013AcA....63..405G} due to dust (D-type). D-type Sb are known to have larger orbital periods on the order of at least $>$20 years. Based on the position of ZTF18aaypgug on the IPHAS color-color diagram, such a system is likely to have an H$_{\alpha}$ EW emission of at least $\sim$1000 Å. Assuming the D-type Sb classification is a reasonable classification, this would be in line with why our LS periodogram attempts were unable to recover any meaningful period constraints from the relatively short baseline of such unique systems. We conclude that while ZTF18aaypgug is not our targeted main-sequence dipper candidate, we encourage follow-up and source characterization.

\section{Discussion} \label{sec:discussion}

\subsection{Classification Scenarios}
In the following sections, we examine potential classification scenarios for our identified population on a case-by-case basis, considering common distinguishing characteristics and underlying physical mechanisms.

\subsubsection{Eclipsing Binaries} \label{sec:eb}

In this section, we examine the possibility that some fraction of our light curve sample may be canonical eclipsing binaries (EB) as potential sources of contamination. Stellar multiplicity is ubiquitous across both low- and high-mass stars, with studies incorporating theory and observations suggesting a multiplicity fraction of FGK stars in the order of 40–50\% \citep{2023ASPC..534..275O}. The wide-binary fraction for FGK stars within 100 pc is on the order of 10$\%$ \citep{2021A&A...649A...6G}. 

Our light curve analysis revealed multiple lines of evidence against EB contamination in our main-sequence sample. Most compelling are the distinctive morphological characteristics of our light curves. First, in Section \ref{sec:lc-shape}, we demonstrated that many of our light curves show asymmetric single erratic dimming profiles rather than multiple symmetric occultations expected from EBs. Second, highlighted in Section \ref{sec:period_search}, we do not find any periodic light curves in our sample, except for one. While ZTF's sparse sampling could potentially mask some features, the consistent pattern of asymmetric and non-periodic dimmings argues against the case of canonical EB. 

Next, the timescale of our main-sequence dippers is not compatible with those of typical stellar EBs. To estimate the expected eclipse durations, we assume simple Keplerian circular orbits. While we acknowledge that eccentricity tends to increase with orbital period \citep{2016MNRAS.456.2070T}, which could lead to longer eclipses at periastron, the overall duration of an eclipse is still constrained by the companion’s transverse velocity and primary/secondary radii, given by:
\begin{equation}
    t_{dip} = \frac{2(R_{1} + R_{2})}{v},
\end{equation}
where $R_{1}$ and $R_{2}$ are the radii of the primary and companion, and $v$ is the transverse velocity of the companion across the stellar equator. 

As shown in Figure~\ref{fig:ms_dipper_timescale_depth}, the median timescales of our events range from weeks to months. Assuming a solar-mass primary star with a negligible mass companion in a circular orbit (semi-major axis $\sim$1.8 AU) and a single eclipse within a $\sim$2.5-year ZTF baseline (roughly half of the full ZTF light curve duration), we would expect an eclipse duration of only $\sim$1.5 days (tangential velocity of 21 km/s).On the contrary, producing eclipses lasting weeks to months, comparable to the observed timescales of our main-sequence dippers, would require semimajor axes to be between 10 and 10,000 AU. The probability of a transit occurring for such a wide binary is given by:
\begin{equation}
    P_{transit} \propto \frac{(R_{1} + R_{2}) (1+e)}{\alpha (1-e^2)},
\end{equation}
where $e$ is the eccentricity, $\alpha$ is the semi-major axis, and $R_{1}, R_{2}$ are the radii of the stars. For wide binaries at separations of 10–10,000 AU, the transit probability is on the order of $10^{-4}$ to $10^{-7}$ assuming an eclipse between a solar and sub-solar companion with circular orbit ($e$=0), making such chance alignments remarkably rare. While these low transit probabilities could account for the rarity of such events, they fail to explain the complexity of the observed transit shapes. One possible alternative is the simultaneous transit of multiple stars, but the probability of such a rare alignment is increasingly improbable.  

If such wide binaries were the cause of our observed dips, we would expect a considerable fraction of our sample within 1 kpc to be resolved as binaries with separations greater than 1000 AU due to \textit{Gaia}'s superb $\sim$1-arcsecond angular resolution, though, at epochs of transit, the semimajor axis might not be the largest depending on if this is a bounded orbit. Given the observed light curve morphologies and timescales, it is more plausible that some sources are wide binaries with companions enshrouded by circumstellar material, as we will discuss in later sections. Furthermore, a search through known ZTF eclipsing binary catalogs yielded no matches to our sources. We therefore conclude that the vast majority of our main-sequence dipper candidates are unlikely to be due to canonical eclipsing binary systems.

\vspace{1cm}

\subsubsection{Cataclysmic Variables} \label{sec:cv}

Cataclysmic Variables (CVs) exhibit complex photometric variability across timescales ranging from minutes to years, with these binary systems consisting of a primary white dwarf accreting material from a secondary high-mass donor star \citep{1976ARA&A..14..119R}. The observed variability manifests through multiple mechanisms: periodic optical variations on hour-long timescales corresponding to orbital dynamics, and state transitions in the accretion disk producing high-energy outbursts. Nova-like variables display stochastic large-amplitude variability driven by fluctuations in mass-transfer rates, transitioning between high and low accretion states on year-long timescales \citep{2017Natur.552..210S}, while certain geometric configurations where the accretion disk occults the donor star can generate large-amplitude dips over hours \citep{2015MNRAS.452.1060C}. 

As discussed in Section \ref{sec:aux-survey}, 33 stars were identified in the IPHAS catalog, one of which showed a clear excess in H$\alpha$, while the remaining had H$\alpha$ and broad colors consistent with the MS locus. Many subtypes of CVs display H$\alpha$ emission, which can often be found in photometric H$\alpha$ surveys \citep{2006MNRAS.369..581W} that have equivalent widths upward of 10 Å. Though this eliminates the interpretation for the scenario of those 32 stars, the remaining MS stars could be potential candidates since we do not have H$\alpha$ constraints. 

The broadband optical photometry collated in this study also disfavors the scenario of a CV interpretation. By taking a catalog of known CVs from \citet{2022ApJ...938...46A}, we crossmatched all sources to the PS1 catalog to obtain broadband \textit{grizy} photometry. We then used $\texttt{isochrones}$\footnote{\href{https://isochrones.readthedocs.io}{https://isochrones.readthedocs.io/en/latest}} to simulate the stellar FGKM main-sequence locus for stars between ages of 7-10 Gyr, assuming a uniform metallicity prior and maximum distance of 10 kpc. We used a  Gaussian Kernel Density Estimator (KDE), assuming an adaptive bandwidth using the Silverman approximation \citep{1986desd.book.....S}, to model the shape of the stellar locus. To statistically validate if any given main-sequence dipper color locus is consistent with either the CV or main-sequence locus, we implemented Bayes' theorem:  
\begin{equation}
    P(Class | X) = \frac{P(X|Class) P(Class)}{P(X)},
\end{equation}
where the class is either a main-sequence or cataclysmic variable. The evidence term can be written analytically, assuming that the product of the prior and the likelihood function over all alternative hypotheses: 
\begin{equation}
   P(X) = \sum P(H_i) P(X|H_{i})
\end{equation}

For each of our candidate stars, using its corresponding PS1 \textit{grizy} photometry, we computed the posterior distribution that it belongs to either the main-sequence or CV, assuming a uniform prior. In all 81 cases, no candidate exceeds a 50\% probability of belonging to the CV population. Based on the 4-dimensional color-color locus of the PS1 photometry, we have considered this study. We conclude that all of our sources are consistent with the stellar main-sequence locus and are likely not CVs. In addition, the lack of any detections from the Galaxy Evolution Explorer (GALEX; \citet{2005ApJ...619L...1M}) prevents firm conclusions about the excess UV emission. However, all our candidates have redder and brighter Gaia colors compared to canonical CV's, which typically lie between the main-sequence and the white dwarf cooling sequence in the Gaia CMD \citep{2025PASP..137a4201R}. In future follow-up studies, if any undetected CV did not meet our search criteria, low to medium-resolution spectroscopy can easily reveal telltale spectroscopic features of emission lines that could test our hypothesis and rule out with higher fidelity. 

\subsubsection{Starspots} \label{sec:starspots}

Low-mass cool stars with convective envelopes exhibit dark regions where the local magnetic field suppresses convection, forming starspots that can cover a significant fraction of the stellar surface and yield a few percent change in flux (\citet{2009A&ARv..17..251S}, and references therein). Pre-Main-Sequence (PMS) stars are characterized by rapid rotation and strong surface magnetic activity, leading to heightened chromospheric emission \citep{2019LRSP...16....3T}. We explored whether starspot activity could explain the observed variability in our main-sequence dipper sample and found multiple inconsistencies that argue against this interpretation. While the timescales of our observed dimming events (weeks to months) overlap with typical starspot lifetimes \citep[e.g.][]{2017MNRAS.472.1618G}, the morphology of the light curves appears highly incompatible with rotational modulation of spot groups. The amplitude of the observed dimming events substantially exceeds expectations for starspot-induced variability. Typical starspot contrasts produce photometric variations of only a few percent, well below our observed dimming depths and below the ZTF detection threshold. The light curve amplitudes, morphologies, and timescales do not well match the interpretation of starspots.

\vspace{1cm}

\subsubsection{Intervening ISM} \label{sec:dust-ism}

It has been proposed that a significant fraction of Galactic baryonic dark matter may reside in cold, dense clouds of gas and dust \citep{1995Ap&SS.234...57D, 1996ApJ...472...34G}. Observations from the Submillimetre Common-User Bolometer Array (SCUBA) have identified cool, compact clumps with temperatures of $\sim$10K, suggesting such cold structures could be Galactic in origin \citep{2001MNRAS.323..147L}. \citet{2002MNRAS.332L..29K} proposed that these dense clouds could produce transient dimming events of background stars with timescales ranging from 10 to 1000 days, and extinction levels reaching 5 magnitudes, leading to sharp cutoffs in stellar light curves. These hypothesized clumps bear similarities to one class of solutions proposed by \citet{TabbySolutions} for the dimming events of the Boyajian star, in which localized interstellar dust structures on sub-AU scales in the Interstellar Medium (ISM) obscure a background star. 

A full investigation of this scenario is beyond the scope of this study for several reasons. First, testing the hypothesis that dark interstellar clouds occult our sources requires precise knowledge of the full kinematics for our stars, which is currently unavailable. This explanation would likely only be viable for sources exhibiting a single discrete dimming event, as multiple photometric dips make the scenario highly improbable. \citet{DrakeBlobs} conducted a search for such events across 4.8 million stars and found no viable candidates with masses in the range of within the 10$^{-4}$--10$^{-2}$ M$_{\odot}$ under any extinction exceeding $A_V>$ 0.2 mag. Our observed rate of main-sequence dippers is therefore inconsistent with the expected event rate for occulting dark clouds, which is predicted to be $>$100 events per year per million stars \citep{2002MNRAS.332L..29K}. Finally, we find no evidence that neighboring sources exhibit similar photometric dips, though one could argue that intervening material might be highly localized.

\subsubsection{Young Stellar Objects} \label{sec:YSO}

Our main-sequence dipper stars exhibit variability possibly resembling YSO ``dipper'' stars, which show quasi-periodic or stochastic flux drops lasting days to years. YSO dippers are attributed to dusty disk structures, including wraps, clumps, and accretion streams \citep{2015A&A...577A..11M, 2014AJ....147...82C, 2011ApJ...733...50M, 2003A&A...409..169B}. In the magnetospheric-accretion paradigm, the elevated inner-disk material briefly crosses our line of sight, occulting the stellar photosphere on the timescale of days \citep{Bodman_Dippers_YSO, 1994ApJ...426..669H}. Beyond short-term events, some YSOs show extreme variability with dips $>$2 magnitudes over months to years. The AA Tau 2011 dimming episode was attributed to increased circumstellar disk density beyond 7.7 AU \citep{2013A&A...557A..77B}. These prolonged dimming episodes typically result from non-axisymmetric density waves and scale height perturbations, possibly induced by planetary-mass companions \citep{2012ARA&A..50..211K, 2017ApJ...836..209R}. 

As noted in Section \ref{sec:lc-char}, 75$\%$ of our sample exhibits single isolated dimming episodes over the 5.5-year ZTF baseline, unlike typical YSO dippers, which characteristically show quasi-periodic variability with multiple dips \citep{ZTF-YSO}. In some rare circumstances, such as the case of EPIC 204376071, a $\sim$10Myr system in the Upper Scorpius Association \citep{2019MNRAS.485.2681R}, which exhibited an 80$\%$ dip that lasted for one day. Similarly, a handful of main-sequence dipper stars with prolonged dimming deep events ($>100$ days) pose a light curve resemblance with AA-Tau analogs. It is conceivable that the remaining 25$\%$ of our main-sequence stars with multiple dips could fit with canonical YSO variability. We found that our main-sequence dippers do not share the ZTF $(g-r)$ color properties of YSO studies \citep{ZTF-YSO}, which are dominated by extinction and accretion effects. Instead, Figure \ref{fig:dipper_gr_colors} shows that some of these stars become redder during their dips, indicating sub-micron dust.

YSO classification relies on infrared and H$\alpha$ properties. Classical T-Tauri stars (CTTs) show H$\alpha$ equivalent widths $> 10$ Å and strong IR excess from inner disks, whereas weak-line T-Tauri stars (WTTs) have equivalent widths $< 10$ Å with little excess, indicating WTTs are more evolved \citep{2015A&A...580A..26G}. In Figure \ref{fig:2mass-wise-var}, we show the 2MASS/WISE color-color diagram for our main-sequence dippers and Type I, II, and Transition Disks identified in the Sh2-112 Filamentary Cloud Complex, including their associated photometric boundaries determined by \citet{2014ApJ...791..131K}. We also overlay the unWISE W$_{1}$ Z-score variability at the color-color position of each star, but do not find any obvious trends. The 2MASS-WISE color-color diagram shows that almost all of our main-sequence dipper stars are bluer than the transition region between class II and transitional disks, and the class I disk. Table \ref{tab:viz_table} presents the case of 8 of our main-sequence dippers stars that have been flagged as class II YSO; however, their indicated probability is $\sim$0. The narrow IR occupied region of our main-sequence dippers could be consistent with either the WWT or the main-sequence stellar locus. Similarly, the color-color IR diagram highlighted in Figure \ref{fig:IPHAS_color_color}, shows that for the fraction of sources that have been observed by IPHAS, show no signs of H$\alpha$ excess.  

\begin{figure}
    \centering
    \includegraphics[width=1.0\linewidth]{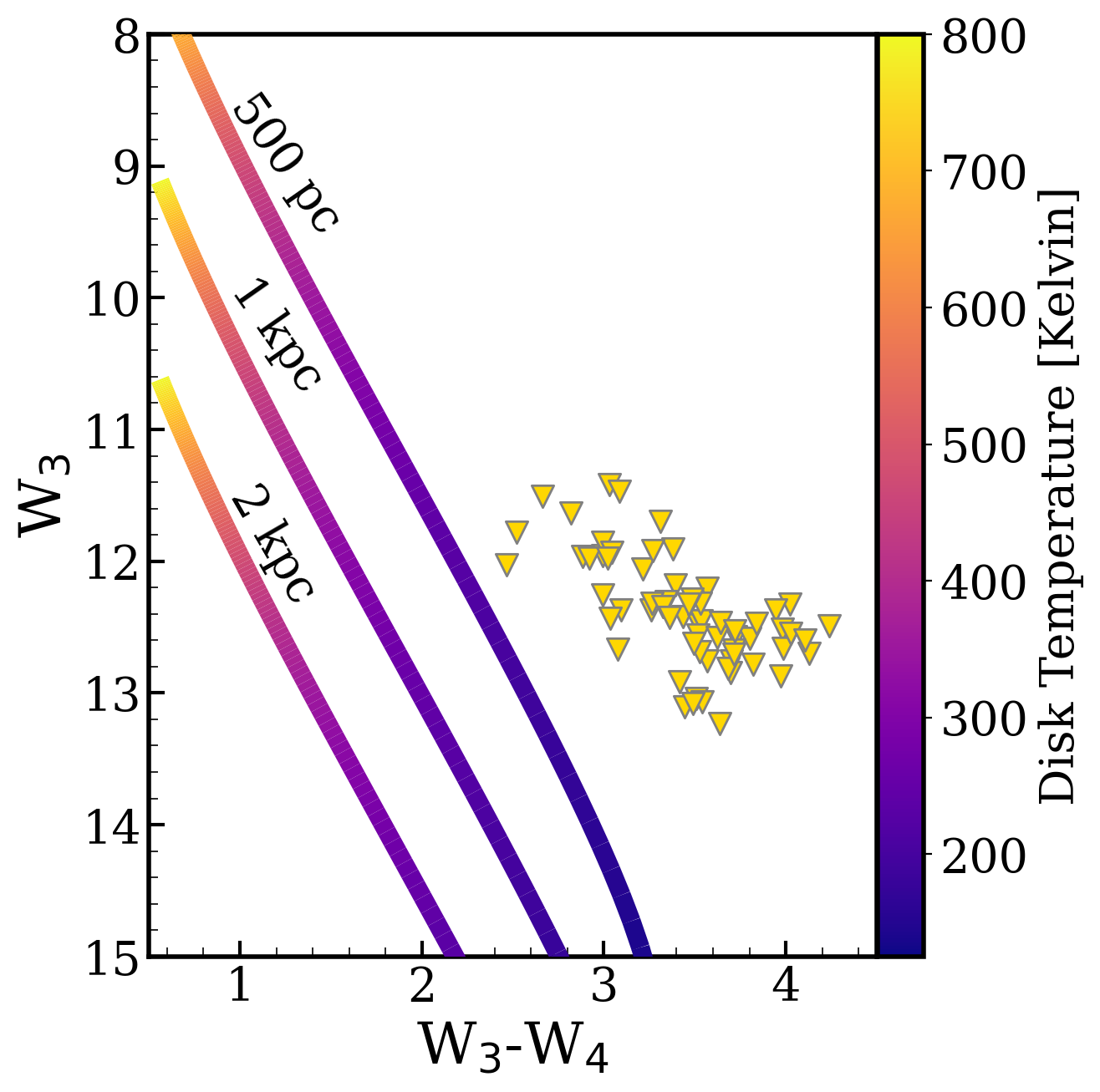}
    \caption{WISE color-magnitude diagram displaying the upper limits of our identified main-sequence dipper stars with SNR $<$2$\sigma$ (gold markers). The lined models represent BT-NextGen SED models, incorporating an additional disk emission component with a radius of 250 AU. The color bar indicates the variation in disk temperature, while the different models are explored across multiple distances.}
    \label{fig:WISE-3-FIR-Models}
\end{figure}

To evaluate the possibility of a disk, in Figure \ref{fig:WISE-3-FIR-Models} we present stellar disk models and compare them to the WISE upper limits. We use the BT-NextGen stellar photospheric model grid \citep{2012RSPTA.370.2765A}, assuming emission from a Sun-like star. Our model includes an excess disk component with temperature (T$_{d}$) and radius (R$_{d}$=250 AU) using \texttt{Species} \citep{Species}. By varying disk temperature and stellar distance, we integrated the spectral energy distribution to obtain theoretical W$_{3}$ and W$_{4}$ bandpasses. In Figure \ref{fig:WISE-3-FIR-Models} we show the color-magnitude diagram of the hypothetical disk excess with varied dust temperatures between 150-800 Kevlin, and distances between 500-2000 pc. Our models predict that the hotter disk $\gtrsim$500 Kelvin at all three distances would most likely preclude being bright enough to be detected with WISE. Based on our models, we find that a 250 AU disk with T$_{s}$$>$500 Kelvin can be eliminated. This can be further argued by the fact that disks of YSO dipper stars near the co-rotational radius of the star are at least $\sim$915 Kelvin (see Table 1 in \citet{Bodman_Dippers_YSO}). Out of the 81 candidates, only two sources have a $>$2 $\sigma$ W$_{3}$ and W$_{4}$ detections. Namely, we have identified one of those stars to be, ZTF18actxeox, which exhibits one $\sim$0.23 mag dip and is located near the H-II LBN 140.77-1.42 and W5 and star-forming regions \citep{2003ApJ...595..880K, 2012A&A...546A..74D} though its distance seems to be closer $\sim$500 pc compared to the estimated 2kpc predicted distance of these H-II regions and its W$_{3}$-W$_{4}$ color has been estimated to be 3.8. The other star has been identified as ZTF19abtaugz with W$_{3}$-W$_{4}$ 3.2; however, we did not find any nearby neighboring sources to determine if it is part of any obvious H-II regions. 

\begin{figure*}
    \centering\includegraphics[width=0.9\linewidth]{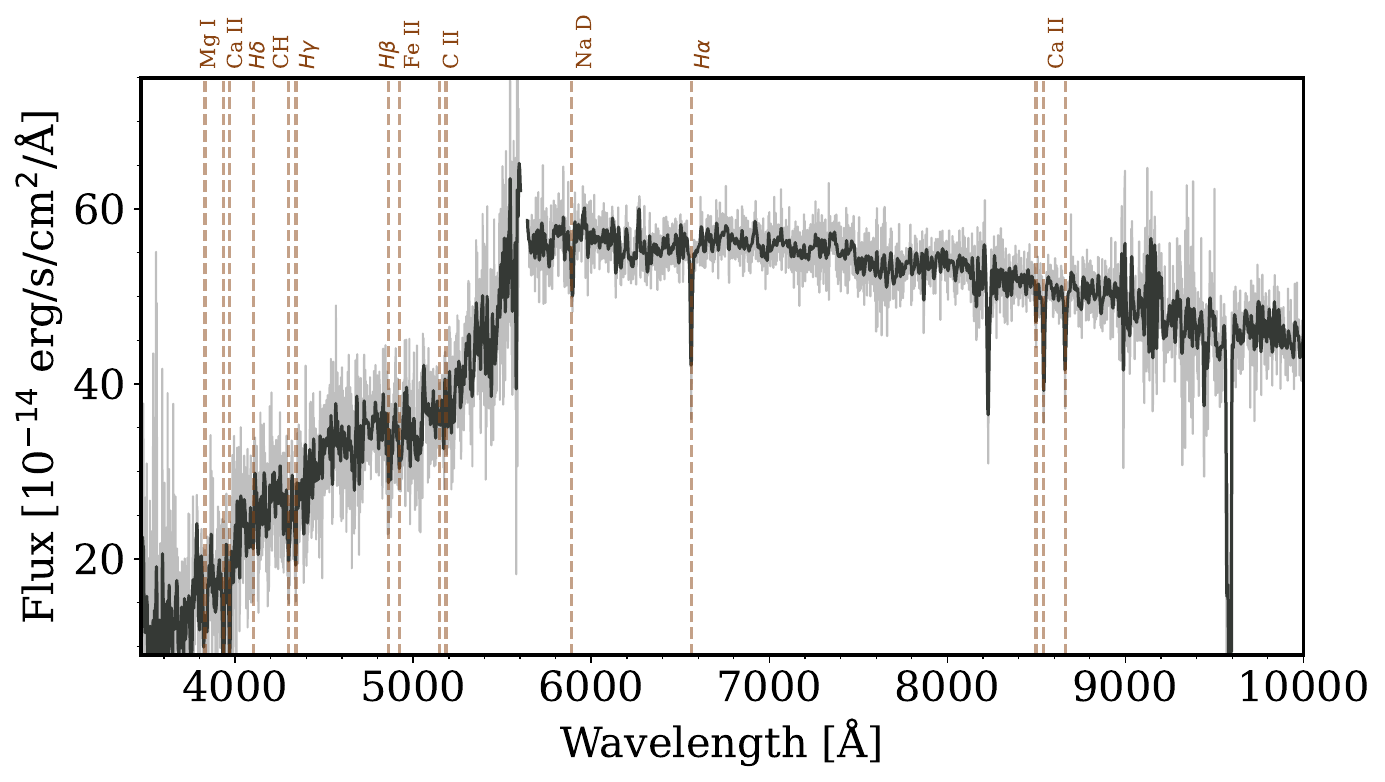}
    \caption{Example of a low-resolution optical spectrum obtained by the P200 for one of our dipper candidates, ZTF18abbtykf. Dashed vertical lines mark the identified absorption features.}
    \label{fig:ZTF18abbtykf_spec}
\end{figure*}

Under the assumption of circular orbits, depths, duration, and lack of periodicity of the dimming events, the transiting material must be located at large separations from the host star, and clump sizes would thus be comparable to the star. However, without deeper IR observations, we cannot rule out that some systems may harbor cool disks responsible for the observed dimmings. Inspection of a low-resolution optical spectrum obtained by the Palomar 200-inch reveals, at least for one of these sources, typical photospheric line FGK absorption lines without the presence of any emission lines, shown in Figure \ref{fig:ZTF18abbtykf_spec}.

\subsubsection{Circumstellar Material} \label{sec:csm} 

Circumstellar disks evolve from primordial gas-rich protoplanetary disks into dust-dominated debris disks as gas dissipates and planets form \citep{DebrisDisksWyatt}. Debris disks are continuously replenished by planetesimal collisions, producing dust observable as IR sources and potentially causing optical stochastic dips \citep{Su_2020, 2022ApJ...927..135S}. Among older low-mass field main-sequence stars with $>$1 Gyr, are found with extreme mid-IR excesses, which may result from frequent planetary collisions \citep{2017AJ....153..165T}.

\begin{figure*}
    \centering
\includegraphics[width=1.\linewidth]{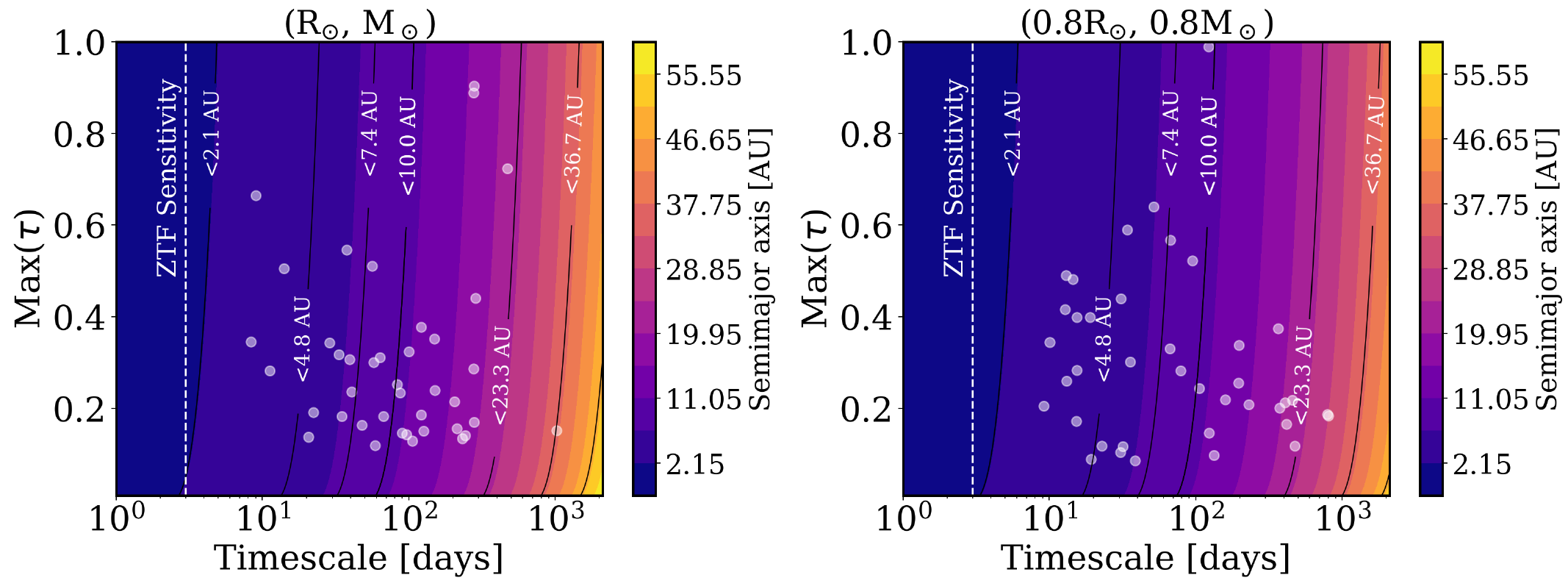}
    \caption{Timescale versus maximum transit depth ($\tau$), defined as $1 -$ minimum normalized flux. The background color contours represent the estimated semimajor axis of the occulting bodies, assuming simple Keplerian circular orbits. The sample is divided into two stellar mass bins: stars with $R_{\odot} \leq 0.8, M_{\odot}$ (right) and those with $R_{\odot} \geq 1, M_{\odot}$ (left). The dashed vertical white line marks the ZTF cadence sensitivity limit. White points indicate the identified dipper candidates.}
\label{fig:timescale_semimajor}
\end{figure*}

The deep and prolonged brightness dips observed in our main-sequence dippers could hint at distant, substantial dust clouds orbiting the star. Assuming this material is bound to the star, we attempt to constrain its semimajor axes. We start with the assumption that the occulting material is completely opaque on a circular equatorial transit, then the fractional depth will be: 
\begin{equation}
    \max(\tau) \approx \frac{2(R_* + S)}{\Delta t},
\end{equation}
where $\tau$ is the fractional normalized transit depth, $\Delta$t is the timescale of the occultation, and the occulting size S. Under this assumption, we can estimate the upper limit to the semi-major axis of such a clump, assuming that it has a mass much less than the primary star and is on a circular orbit with an equatorial transit, then the semimajor axis upper limit will be:
\begin{equation}
    a_{\rm circ} \approx M_* \bigg(\frac{S-R_*}{\Delta t}\bigg)^{-1/2},
\end{equation}
where a$_{circ}$ will be the semimajor axis, M$_*$ will be in solar masses. In Figure \ref{fig:timescale_semimajor}, we separated sources near 0.8 solar radii and masses from the \texttt{StarHorse 2022} parameters, which include a mass estimate derived from the isochrone fitting. In both mass regime scenarios presented in Figure \ref{fig:timescale_semimajor} we note that the median is $\sim$10 AU, which makes sense since the bulk of our dippers have long timescales. We caution against the interpretability of such results and suggest they be understood as upper limits, given competing effects such as observational selection biases (favoring longer-timescale events), circular orbits, and the assumption that the transiting material is optically thick. The presence of transiting dust clumps at large semimajor axes is compelling, as it naturally explains the complex light curve morphology, extended timescales, and lack of infrared excess observed in many main-sequence dippers. This interpretation fits well with the data at our disposal, namely, that collisional evolution of planetesimals at $\geq$ 10 AU can generate long-lived, asymmetric dust structures without necessarily producing detectable near-IR signatures (\citealp{WyattDD_Planets} and references therein). As suggested in our light curve shape analysis in Section \ref{sec:lc-shape}, many of our discoveries have clear asymmetric shapes that can support the theory of transiting dust structures. It is plausible that structures that persist over Myr timescales provide a mechanism for transient dimming events in systems where circumstellar material periodically obscures the star on such large timescales. We also note in Section \ref{sec:demographics} that at least three of our main-sequence candidates appear to be in open clusters, are appropriate ages where planetesimal collision and formation are expected, and thus can be the origin of the circumstellar material. Interestingly, material near $\sim$2 AU under Keplerian velocity would have an orbital period on the order of $\sim$3 years, assuming negligible mass compared to its corresponding solar mass, which would suggest that with the 5.5-year-long ZTF baseline to see at least one more transit of such occulting material.

\subsubsection{Disk Occultation} \label{sec:disk-occultation}

Circumstellar and circumplanetary disks represent important evolutionary stages in stellar and planetary systems, and are considered a ubiquitous feature among young stars \citep{Hughes_review}. Stellar eclipses by disks provide rare opportunities to study their structure, composition, and dynamics. The transits of such disks around their primary stars are thought to be rare. For example, it has been found that the fraction of detached EBs that host disks in the Large Magellanic Cloud (LMC) is approximately 1/1000, and mostly biased toward young systems \citep{2014MNRAS.441.3733M}, while \citet{2012AJ....143...72M} estimated that transiting events around a weak-line T-Tauri star with a secondary low mass companion with a disk to be at least 1/100.

In a handful of instances, there have been reports of transiting disks around young stars \citep{2024A&A...688L..11P, 2024AJ....167...85H, 2015ApJ...800..126K, 2019MNRAS.485.2681R} that could be intriguing laboratories for understanding disk structures and perhaps the formation sites of disk formation around stars, exomoon formation \citep{2015ApJ...800..126K} and planetary interior structure \citep{2011ApJ...734..117S}. However, these confirmed candidates exhibit a diverse range of light curve morphologies, timescales, ages, and color properties that complicate the identification of disk occultation events solely on photometry. Some events show highly symmetric profiles, while others display complex dimming events with multiple minima that can be attributed to disk substructure \citep{2015ApJ...800..126K}. Asymmetric light curve ingress and egress profiles have also been reported, potentially resulting from projection effects of inclined disks creating elliptical occulters \citep{beyonce}.

As discussed in Section~\ref{sec:csm}, due to the large timescales and depths of our observed dips, the occulting material must be at large semimajor axes. If we interpret the results from Figure \ref{fig:timescale_semimajor} such that many of our systems have large semimajor axes, then the Hill sphere of a 1 M$_{\oplus}$ orbiting a sun-like star would have a $\sim$21.5 $R_\odot$ hill radius at $\sim5$AU, well enough to replicate the occultation size ratios observed in some instances of this sample though at the moment the formation mechanisms of such disks remain unclear. 

Detailed radial velocity follow-up is needed to confirm the presence of another body that could contain such rings or disks. Assuming circular orbits, periods must be on the order of $>$5.5 years to explain the lack of recurring transits.

\section{Conclusion} \label{sec:conclusion}

This work presents the results from the first comprehensive search for main-sequence dippers among 63 million FGK main-sequence stars using ZTF Zubercal photometry. Our statistical light curve scoring framework has successfully identified 81 new dipper candidates exhibiting a diverse set of photometric behaviors, some reminiscent of KIC 8462852, others, such as those seen among occultations by disks. Our findings represent a significant expansion of the known population of main-sequence stars experiencing dimming episodes, some of which we hypothesize to be true analogs of Boyajian's star, and analogous stellar systems being occulted by dusty clumps of circumstellar material whose origin remains undetermined. Our conclusions are as follows: 

\begin{itemize}
    \item We developed and implemented a novel statistical light curve scoring framework, enabling efficient analysis of 63 million stellar light curves. Our novel search has revealed 81 new dipper candidates. 76.5$\%$ of these candidates have only one dip within the ZTF photometric baseline of 5.5 years, while the remaining have fewer than 23.5$\%$ $<$3 dips. Due to systematic errors present in the ZTF data, visual inspection is still fundamental in identifying bona fide candidates.
    
    \item Our sample reveals a large spread in dipper light curve morphologies, with timescales spanning 3 days to $\sim$3 years, and depth amplitudes spanning from 1-99$\%$. We find a marginally statistically significant excess of main-sequence dippers with asymmetric dimmings, with a slight excess with longer egress profiles, which is highly compatible with the scenario of transiting circumstellar dust debris.

    \item We investigated the ZTF $(g-r)$ color evolution across the light curves of our main-sequence dippers and found a spread of colors, though marginally, with more events that typically reddened by 0.2 mag. Through an analysis of the ZTF-$gr$ CMD, we found their slopes to be below the typical grain sizes of the ISM, possibly pointing to either smaller particle size or different particle properties. We find several statistically significant correlations, such as dip amplitude and $(g-r)$ color, and timescale versus the ZTF-$gr$ CMD slope. 

    \item An analysis of auxiliary infrared and H$\alpha$ photometric datasets was conducted. The temporal variability of WISE W$_{1,2}$ data did not reveal any variability beyond a $>3\sigma$ level on a $\sim$4 month timescale. The position of the 2MASS-WISE and IPHAS CMD indicates nominal positions where the stellar main-sequence locus, suggesting that the majority of our stars have no infrared excess within wavelengths $<4\mu \text{m}$, and no H$\alpha$ emission.

    \item At least 4 of our main-sequence dipper stars are coincident and members of known open clusters and star-forming regions at ages consistent with planetary impacts. Based on their absolute scale height distributions, we propose that most stars are consistent with a low-alpha Galactic disk, indicating their relative youth.
    
    \item Based on the dip timescale, depth, frequency, lack of periodicity, and infrared excess, we propose that the majority of discovered main-sequence dipper stars are the result of transiting circumstellar dust clumps. Under the assumption of circular orbits, the observed timescales suggest typical semimajor axes of $\sim$5-10 AU. If true, such main-sequence dipper stars may be one-of-a-kind laboratories for understanding the origins of circumstellar dust features at large semimajor axes, intimately tied to planet formation and evolution. 
\end{itemize}

Given the discovery of these puzzling dipper stars, it becomes evident that more auxiliary data is necessary to perform a more robust classification and remove any potential outliers. Namely, deeper IR imaging and high-resolution R$\sim$10000 Å spectroscopy will be required to provide more concrete clues to robust classification scenarios. The upcoming Vera C. Rubin Observatory Legacy Survey of Space and Time \citep[LSST;][]{2019ApJ...873..111I} will provide crucial measurements through high-precision photometry across the \textit{ugrizy} bandpasses. Additionally, the Nancy Grace Roman Space Telescope and follow-up observations from \textit{JWST} will extend our sensitivity beyond 1 $\mu$m, potentially revealing new populations of such systems and providing critical constraints on the nature of the occulting material.


\section{Acknowledgments}

We are grateful to the anonymous referee for useful comments during the review process. We thank Ann Marie Cody, Andrew Drake, Tobin Wainer, and Ishan Ghosh-Coutinho for their insightful discussions related to this paper. Additionally, we thank Christoffer Fremling for obtaining spectra for us using the Palomar 200-inch telescope. AT and JRAD acknowledge support from Breakthrough Listen. The Breakthrough Prize Foundation funds the Breakthrough Initiatives, which manage Breakthrough Listen. AT acknowledges the generosity of Eric and Wendy Schmidt, by recommendation of the Schmidt Futures program through the LSST Discovery Alliance (LSST-DA), is building an LSST Interdisciplinary Network for Collaboration and Computing (LINCC) program. LINCC Frameworks is supported by Schmidt Sciences. A.T. also acknowledges the support from the DiRAC Institute in the Department of Astronomy at the University of Washington. The DiRAC Institute is supported through generous gifts from the Charles and Lisa Simonyi Fund for Arts and Sciences and the Washington Research Foundation. This publication makes use of data products from the Wide-field Infrared Survey Explorer and NASA/IPAC Infrared Science Archive, which is a joint project of the University of California, Los Angeles, and the Jet Propulsion Laboratory/California Institute of Technology, funded by the National Aeronautics and Space Administration.

\software{\texttt{Astropy} \citep{astropy:2013, astropy:2018, astropy:2022}, \texttt{IPython} \citep{PER-GRA:2007}, \texttt{Matplotlib} \citep{Hunter:2007}, \texttt{NumPy} \citep{harris2020array}, \texttt{SciPy} \citep{2020SciPy-NMeth}}

\newpage

\bibliography{sample631.bbl}{}
\bibliographystyle{aasjournal}

\clearpage
\appendix
\section*{Appendix A} \label{ape:appendix1A}

Here we present a compilation of all our identified main-sequence dipper ZTF light curves in the $g$ and $r$ bandpasses. The data have been collated from IRSA Data Release 22. Figures \ref{fig:comp_lc_single_1}-\ref{fig:comp_lc_single_4} show light curves exhibiting a single identified dipping event, while Figures \ref{fig:comp_lc_multi_1}-\ref{fig:comp_lc_multi_2} display systems with multiple dimming events detected during our observational baseline. At the top of each light curve, we list the ZTF coordinates for each identified star. 

\vspace{1em} 

\begin{figure}[H]
  \centering
  \includegraphics[width=0.9\textwidth]{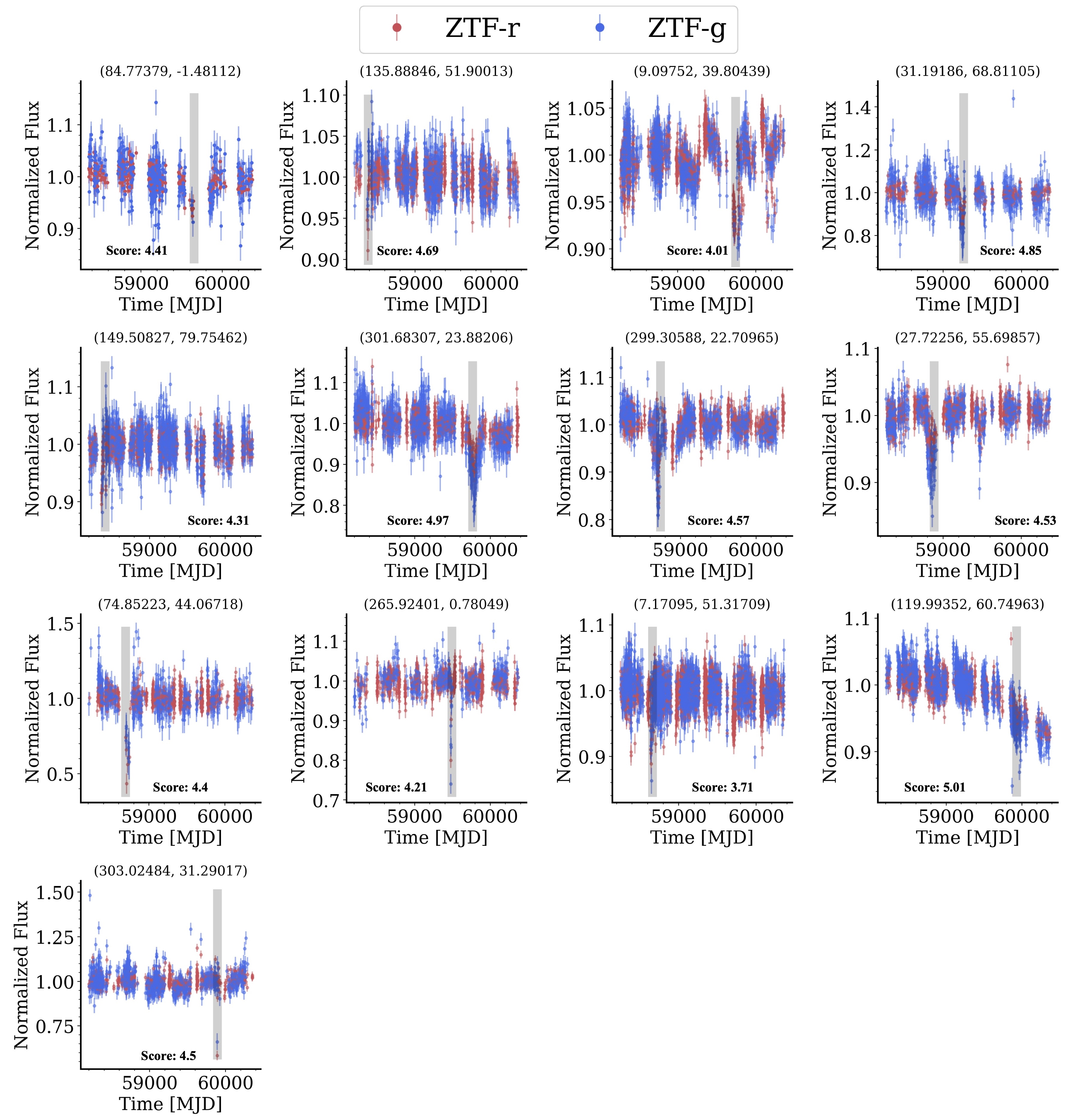}
  \caption{Normalized flux ZTF $gr$ light curves identified with one unique dip. The black shaded line indicates the time of the photometric dip identified in this study.}
  \label{fig:comp_lc_single_4}
\end{figure}

\newpage
\begin{table*}
\centering 
    \begin{tabular}{c c c c c c c c}
    \hline 
    Gaia DR3 ID & R.A & Dec & G$_0$ \footnote{\texttt{StarHorse 2022} absolute magnitude corrected from extinction. } & G$_{BP}$ - G$_{RP}$\footnote{\texttt{StarHorse 2022} dereddened BP-RP Gaia color.} & Distance\footnote{50$^{ith}$ percentile distances derived from \texttt{StarHorse 2022.}} & t$_{\text{dip}}$\footnote{Mean time of the deepest dip.} & N$_{\text{Dips}}$ \\
    &  &  & (mag) & (mag)& (kpc) &(MJD) & \\
    \hline\hline 
4319330107178019200 &  288.43875 & 13.36944 & 4.288 & 0.678 & 2.091 & 58643.299 & 1.0 \\
3121854222422204544 &  93.59962 & -0.25528 & 7.026 & 1.243 & 0.86 & 58787.517 & 1.0 \\
446400591305341056 &  49.02593 & 54.09025 & 3.696 & 0.781 & 1.104 & 59055.467 & 1.0 \\
2285460370033812864 &  327.52656 & 80.14685 & 5.313 & 1.01 & 1.274 & 59779.271 & 1.0 \\
198430753251200128 &  73.67696 & 37.70347 & 4.931 & 0.838 & 1.084 & 58540.282 & 1.0 \\
2082534881979041280 &  303.76791 & 46.45553 & 5.867 & 1.101 & 0.33 & 58252.489 & 1.0 \\
2079894366138598656 &  294.41874 & 44.81552 & 4.957 & 0.79 & 4.301 & 59484.241 & 1.0 \\
2174227726473890176 &  324.54651 & 53.85427 & 5.047 & 0.865 & 0.899 & 58799.222 & 1.0 \\
428654920313727744 &  7.68843 & 60.64189 & 5.561 & 0.929 & 1.851 & 58333.34 & 1.0 \\
508972801614364032 &  24.37886 & 58.27367 & 4.658 & 0.824 & 1.793 & 58877.125 & 1.0 \\
512103523536262144 &  25.53477 & 63.66452 & 4.328 & 0.844 & 0.643 & 59065.39 & 1.0 \\
2126217688671428864 &  290.79236 & 44.36126 & 4.734 & 0.796 & 0.987 & 58257.476 & 1.0 \\
1989462803234296832 &  345.37488 & 52.66552 & 6.91 & 1.361 & 0.496 & 59816.387 & 1.0 \\
4208785414254570368 &  295.93436 & -6.18203 & 4.582 & 0.791 & 1.673 & 58318.334 & 1.0 \\
4165810383809607808 &  267.36017 & -7.41839 & 3.868 & 0.765 & 1.473 & 58291.272 & 1.0 \\
166212073259272832 &  65.34766 & 30.73762 & 5.994 & 1.083 & 0.432 & 59922.367 & 1.0 \\
2686018434626245888 &  321.69203 & -1.25194 & 5.269 & 1.064 & 1.145 & 58326.377 & 1.0 \\
2168271294020636800 &  311.1466 & 49.98304 & 4.31 & 0.806 & 1.581 & 58271.453 & 1.0 \\
1956436050910998784 &  333.13294 & 40.09462 & 6.713 & 1.303 & 0.878 & 59126.323 & 1.0 \\
2067746141339717120 &  307.91981 & 40.76595 & 3.941 & 0.701 & 1.295 & 58646.477 & 1.0 \\
1862110906882967552 &  306.6495 & 31.67113 & 6.091 & 1.11 & 1.069 & 59525.134 & 1.0 \\
2279847496317770240 &  339.21377 & 75.91337 & 5.282 & 0.911 & 2.673 & 58771.335 & 1.0 \\
425286325927143936 &  11.79256 & 59.35174 & 4.606 & 0.775 & 1.43 & 59453.343 & 1.0 \\
184965210441328896 &  75.62383 & 33.96783 & 4.972 & 0.89 & 1.414 & 59426.487 & 1.0 \\
200936780769225088 &  75.61856 & 41.33001 & 6.059 & 1.016 & 0.832 & 59542.202 & 1.0 \\
... & ... & ... & ... & ... & ... & ... & ... \\
1132318104107576064 &  149.50826 & 79.7546 & 6.564 & 1.218 & 0.865 & 58364.507 & 2.0 \\
5677433705699331328 &  144.75644 & -20.05037 & 5.684 & 0.935 & 1.231 & 59198.439 & 2.0 \\
4898717140928169344 &  69.11309 & -21.21456 & 6.502 & 1.067 & 2.094 & 59260.149 & 2.0 \\
2072562856964901760 &  299.20706 & 38.80473 & 5.698 & 1.002 & 2.231 & 59689.447 & 2.0 \\
    \hline 
    \end{tabular}
\caption{Main sequence dipper candidates including their observed properties. This table will be provided in its entirety in machine-readable form in the online journal. \label{table:final_candidate_table}} 
\end{table*}

\



\begin{figure*}
\includegraphics[width=0.93\textwidth]{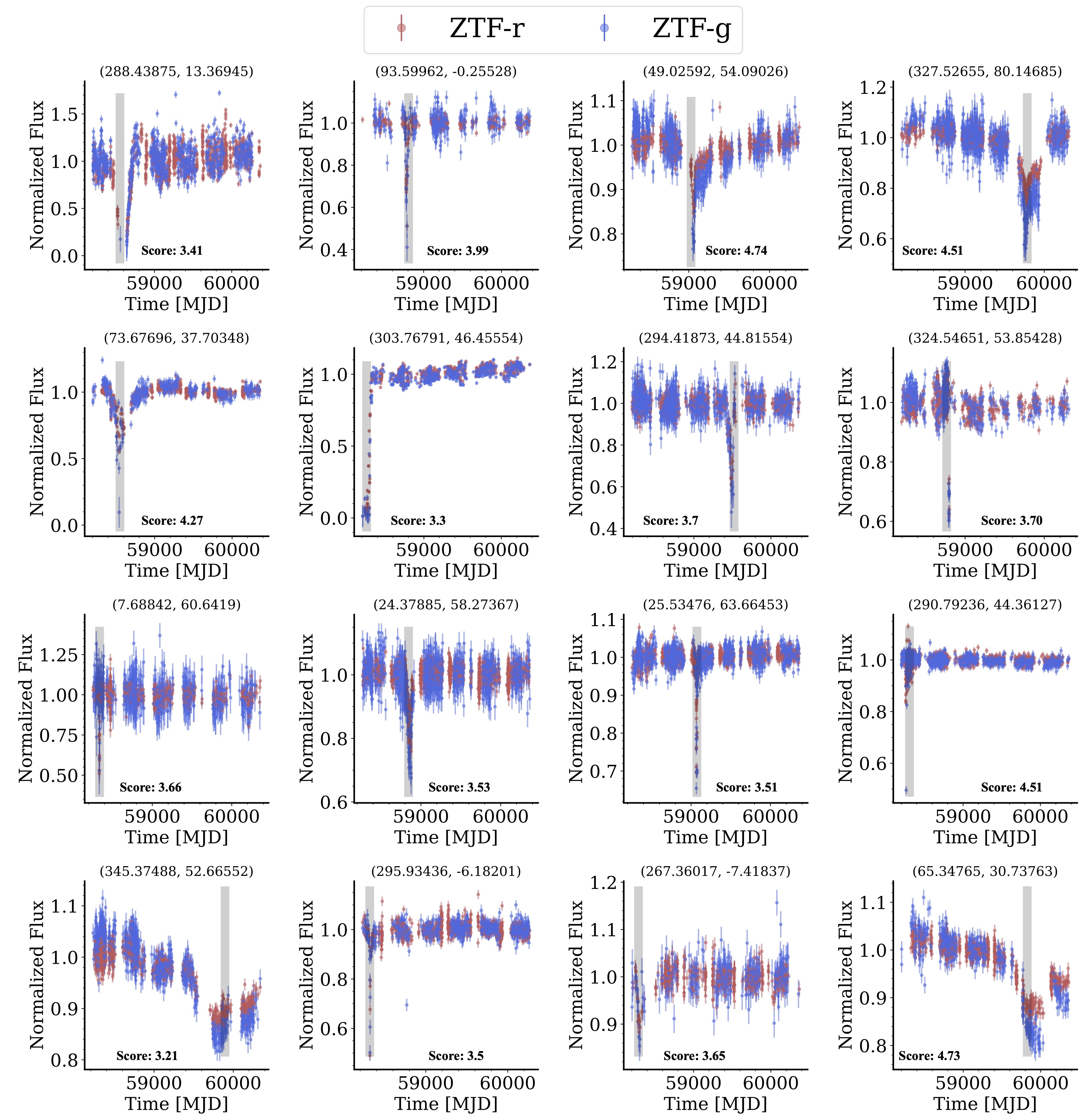}
  \centering 
  \caption{Continued, normalized flux ZTF $gr$ light curves identified with one unique dip. The black shaded line indicates the time of the photometric dip identified in this study.}
  \label{fig:comp_lc_single_1}
\end{figure*}

\newpage
\begin{figure*}
\includegraphics[width=0.93\textwidth]{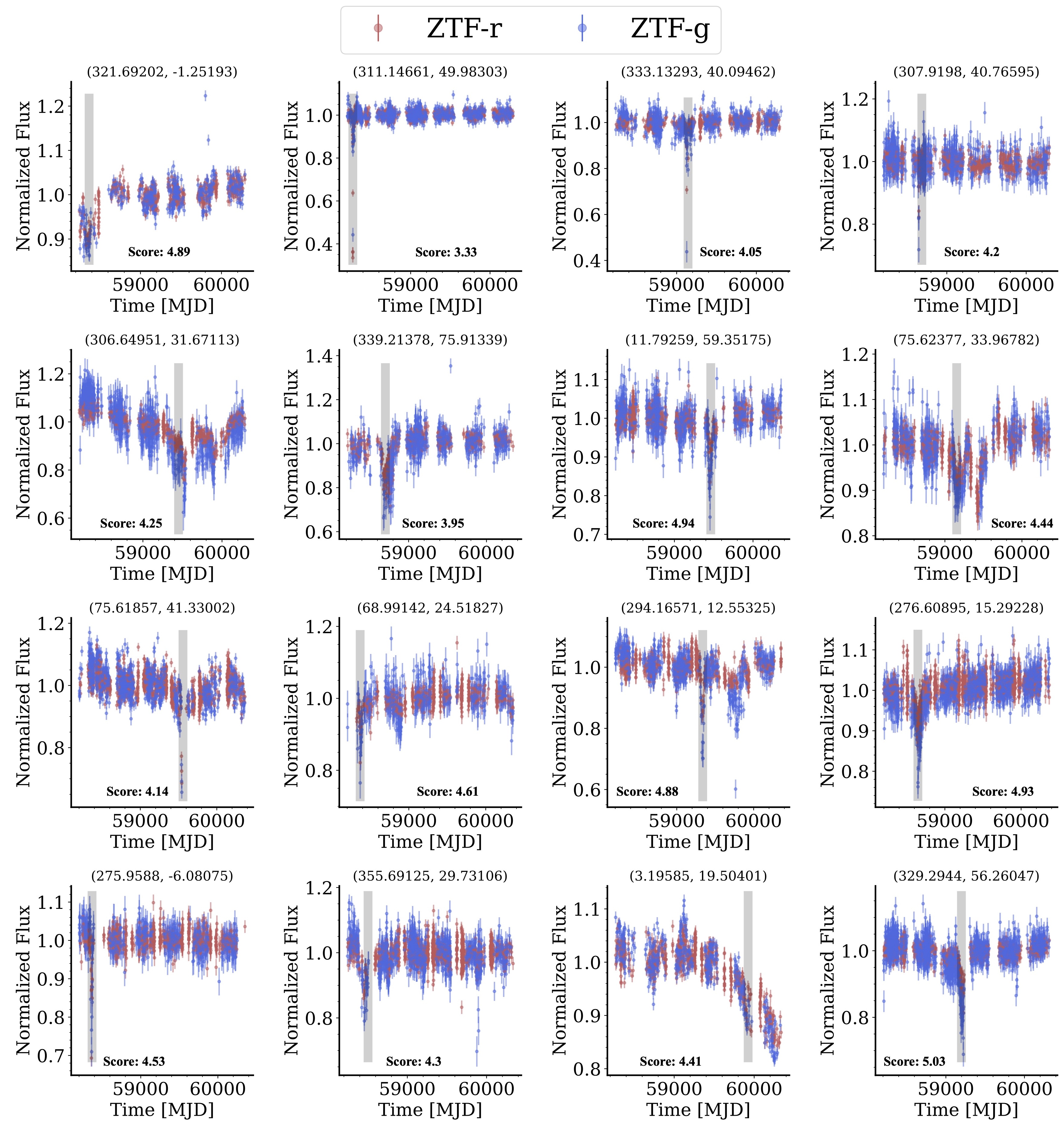}
  \centering 
  \caption{Continued, normalized flux ZTF $gr$ light curves identified with one unique dip. The black shaded line indicates the time of the photometric dip identified in this study.}
  \label{fig:comp_lc_single_2}
\end{figure*}

\newpage
\begin{figure*}
\includegraphics[width=0.93\textwidth]{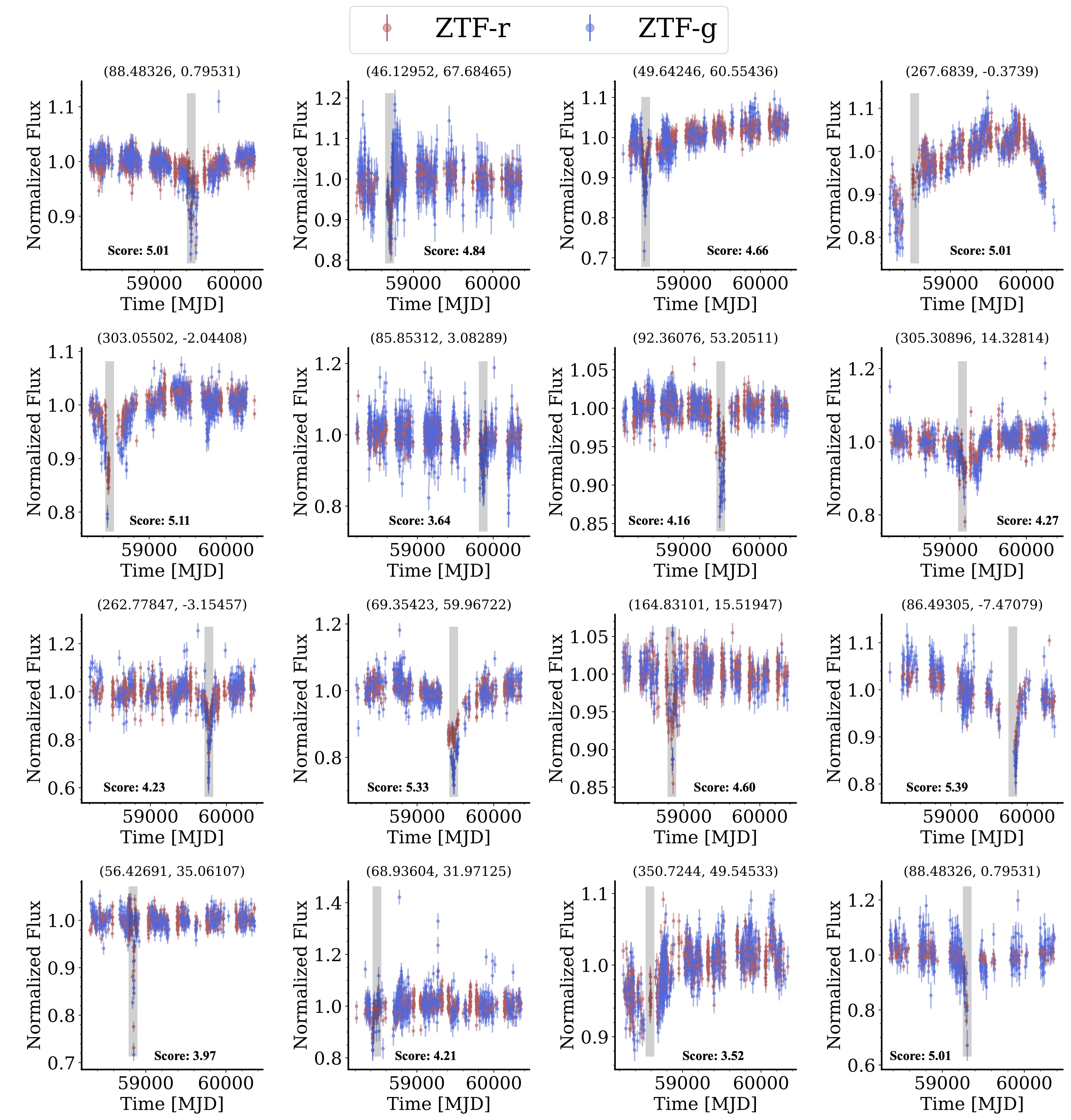}
  \centering 
  \caption{Continued, normalized flux ZTF $gr$ light curves identified with one unique dip. The black shaded line indicates the time of the photometric dip identified in this study.}
  \label{fig:comp_lc_single_3}
\end{figure*}

\newpage
\begin{figure*}
\includegraphics[width=0.93\textwidth]{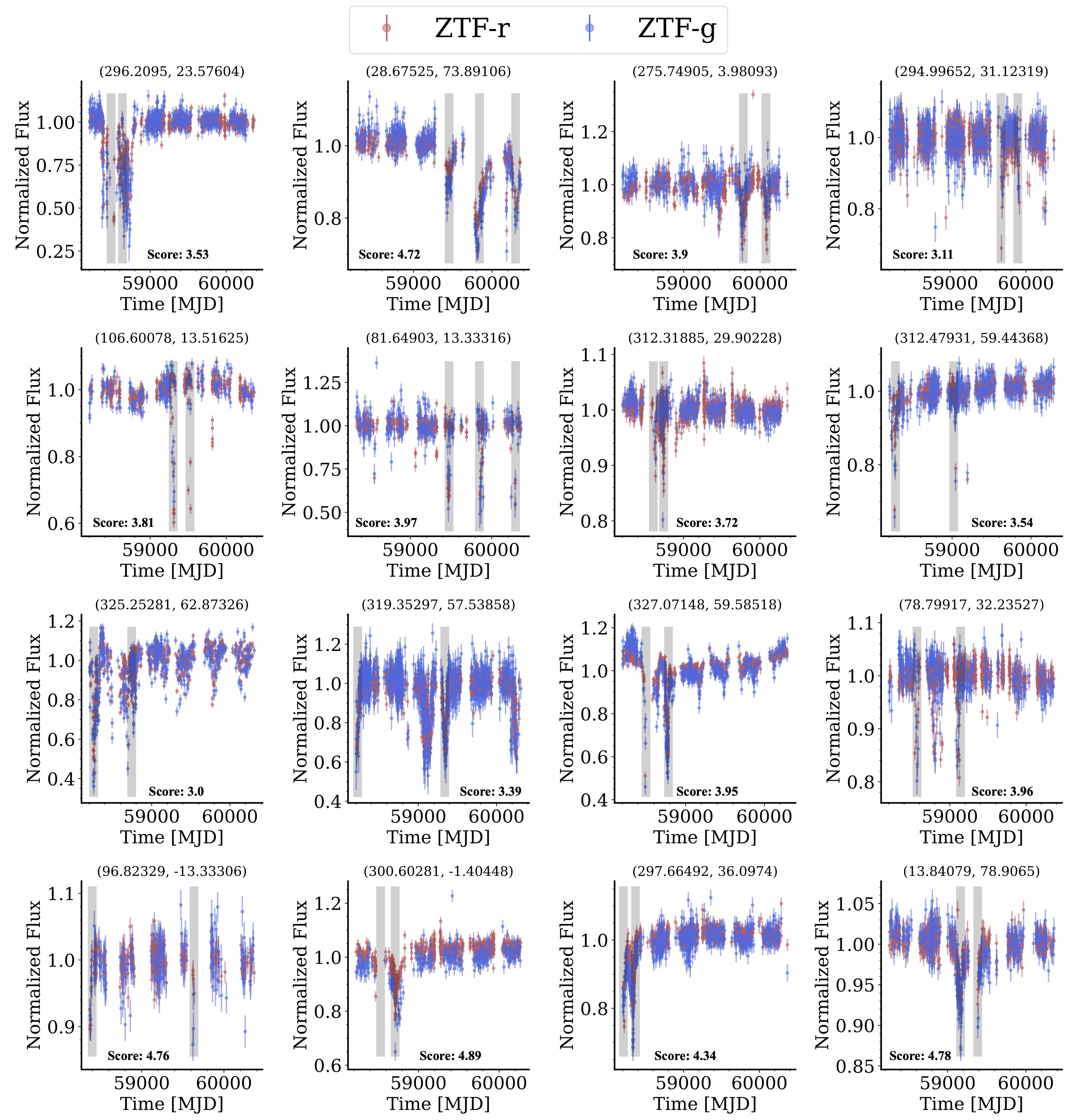}
  \centering 
  \caption{Normalized flux ZTF $gr$ light curves identified with multiple dips. The black shaded line indicates the time of the photometric dip identified in this study.}
  \label{fig:comp_lc_multi_1}
\end{figure*}

\newpage
\begin{figure*}
\includegraphics[width=0.93\textwidth]{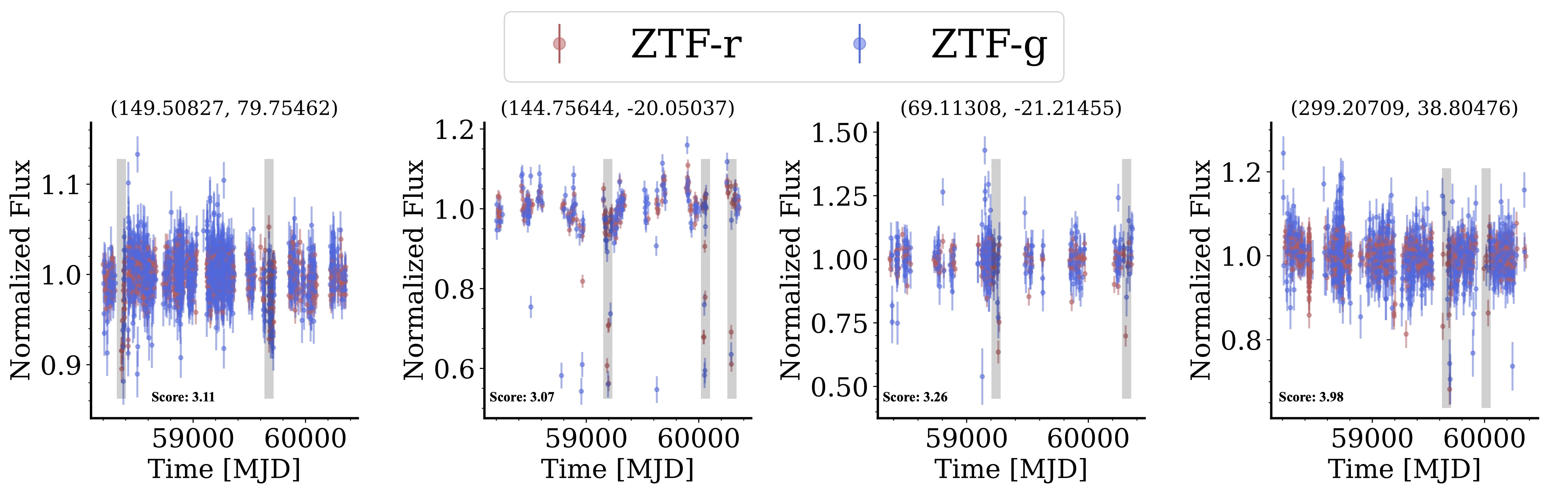}
  \centering 
  \caption{Continued, normalized flux ZTF $gr$ light curves identified with multiple dips. The black shaded line indicates the time of the photometric dip identified in this study.}
  \label{fig:comp_lc_multi_2}
\end{figure*}

\clearpage
\section*{Appendix B} \label{ape:appendix1B}

\begin{figure}[H]
    \centering
    \includegraphics[width=0.5\linewidth]{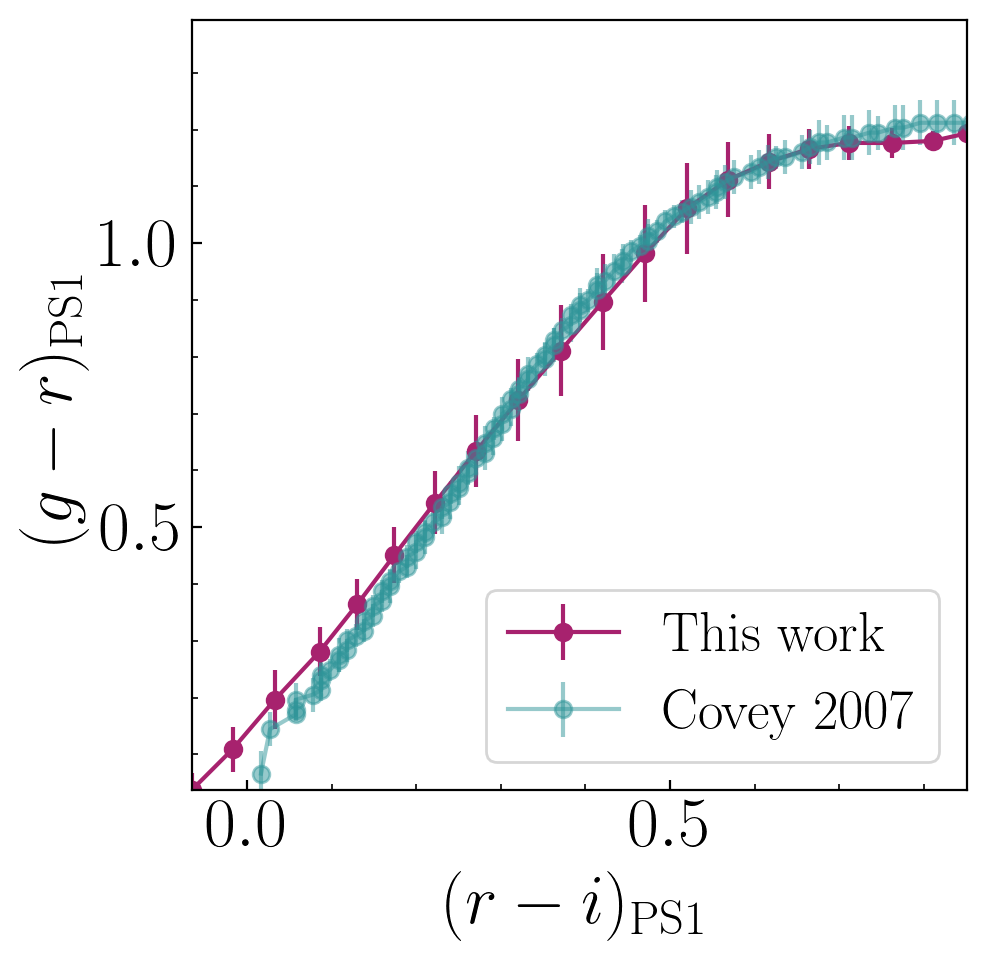}
    \caption{Comparison of the Pan-STARRS1 color-color diagram between this work and \citet{2007AJ....134.2398C}.}
    \label{fig:locus_ps1_comparison}
\end{figure}

For direct comparison to our final stellar locus selection, in Figure \ref{fig:locus_ps1_comparison} we present the Pan-STARRS1 color-color $(r-i)$ versus $(g-r)$ diagram compared to the work of \citet{2007AJ....134.2398C} and Table \ref{table:locus_ps1_tbl} with our color-color stellar locus values. Since the original color values of \citet{2007AJ....134.2398C} were reported in the SDSS filters, we applied the appropriate photometric corrections to convert them to the PS1 system \citep{2012ApJ...750...99T}.

\begin{table}[ht]
\centering
\begin{tabular}{cc}
\hline
$(r - i)_{\textit{PS1}}$ & $(g - r)_{PS1}$ \\
\hline
-0.03 & 0.10 \\
0.07  & 0.25 \\
0.14  & 0.39 \\
0.21  & 0.53 \\
0.30  & 0.68 \\
0.39  & 0.84 \\
0.48  & 0.99 \\
0.56  & 1.11 \\
0.64  & 1.16 \\
0.72  & 1.18 \\
0.81  & 1.18 \\
\hline
\hline
\end{tabular}
\caption{Pan-STARRS1 color-color locus presented in this work.}
\label{table:locus_ps1_tbl}
\end{table}

\clearpage
\section*{Appendix C} \label{ape:appendix1C}

We conduct a 1-arcsecond cone search in VizieR around every main-sequence dipper candidate in search of potential classification of contaminating sources. Table \ref{tab:viz_table} presents all objects returned that have been potentially flagged to belong to another variable star classification (e.g., \ YSO, AGN, RR Lyrae, BYDRa) or belong to a star-forming and open cluster. For each identification, we provide a brief comment on the classification of the source. We note that these sources remained in the working sample, but we mark them here for cautious interpretation of true main-sequence dippers.

\newcommand{\wrapc}[1]{\vtop{\hsize=9cm \raggedright #1\par}}

\begin{deluxetable*}{cccc}[ht]
\tablecaption{Summary table of potential contaminant candidates.\label{tab:viz_table}}
\tablehead{
  \colhead{RA (deg)} &
  \colhead{Dec (deg)} &
  \colhead{Reference} &
  \colhead{Comments}
}
\startdata
288.43875 & $+13.36945$ & \citep{2023ApJS..266...34G} & \wrapc{Possible planetary nebulae. Marked with H$\alpha$ excess emission discussed in Section~\ref{sec:aux-survey}.} \\
49.02592  & $+54.09026$ & \citep{2023MNRAS.521..354W} & \wrapc{Flagged as potential class~II YSO. However, the posterior probability of belonging to this class is very close to 0.} \\
327.52655 & $+80.14685$ & \citep{2014MNRAS.443..725G} & \wrapc{Flagged as potential AGN, though Gaia astrometry indicates the source is stellar.} \\
73.67696  & $+37.70348$ & \citep{2023MNRAS.521..354W, Kos} & \wrapc{Flagged as potential class II YSO (posterior probability~$\simeq0$). Possibly associated with the open cluster COIN-Gaia-15 with 12\% membership probability.} \\
303.76791 & $+46.45554$ & \citep{2023MNRAS.522...29G} & \wrapc{Flagged as potential ellipsoidal binary candidate with a 3.07 day period; quality score is zero.} \\
324.54651 & $+53.85428$ & \citep{2022ApJ...932..118C} & \wrapc{Flagged as BYDra like variable with score 0.3.} \\
7.68842   & $+60.64190$ & \citep{Groeningen} & \wrapc{Flagged as member of open cluster NGC~129 with 0.1 membership probability.} \\
24.37885  & $+58.27367$ & \citep{Vioque} & \wrapc{Flagged as Herbig Ae/Be star with 0.96 probability of belonging to another class.} \\
25.53476  & $+63.66453$ & \citep{2023MNRAS.521..354W} & \wrapc{Flagged as potential class II YSO; posterior probability $\simeq0$.} \\
319.35297 & $+57.53858$ & \citep{2017AJ....153..204S} & \wrapc{Flagged as potential RR Lyrae class with $\sim$0 RRab/RRc classification score.} \\
321.69202 & $-01.25193$ & \citep{Gavras, 2009MNRAS.398.1757W} & \wrapc{Flagged as BY~Dra like variable and RR Lyrae with `no valid period/score.} \\
333.13293 & $+40.09462$ & \citep{2006SASS...25...47W} & \wrapc{Flagged as rotational variable in the Variable Star IndeX.} \\
307.91980 & $+40.76595$ & \citep{2021MNRAS.508.3388G, 2023MNRAS.521..354W} & \wrapc{Near the Cygnus OB2 complex (Subaru/HSC). Also flagged as potential class II YSO (posterior probability~$\simeq0$).} \\
306.64951 & $+31.67113$ & \citep{2023MNRAS.521..354W} & \wrapc{Flagged as potential class II YSO; posterior probability $\simeq0$.} \\
11.79259  & $+59.35175$ & \citep{2023MNRAS.521..354W} & \wrapc{Flagged as potential class II YSO; posterior probability $\simeq0$.} \\
75.61857  & $+41.33002$ & \citep{2022ApJ...932..118C} & \wrapc{Flagged as BYDra like variable with score 0.2.} \\
78.79917  & $+32.23527$ & \citep{2023MNRAS.521..354W} & \wrapc{Flagged as potential class II YSO; posterior probability $\simeq0$.} \\
3.19585   & $+19.50401$ & \citep{2019RAA....19...64Q, 2024ApJS..272...40Z} & \wrapc{Spectroscopic binary and variable star with 22\,km\,s$^{-1}$ RV amplitude; no validity score. Possible active star (reported $S$--index).} \\
329.29440 & $+56.26047$ & \citep{Kos} & \wrapc{Possibly associated with open cluster UBC 155 (16\% membership probability).} \\
49.64246  & $+60.55436$ & \citep{Vioque} & \wrapc{Flagged as Herbig Ae/Be star with 0.97 probability of belonging to another class.} \\
267.68390 & $-00.37390$ & \citep{2006SASS...25...47W} & \wrapc{Flagged as rotational variable in the Variable Star IndeX.} \\
13.84079  & $+78.90650$ & \citep{2023AJ....166..193J} & \wrapc{Reported counts from LOFAR.} \\
350.72440 & $+49.54533$ & \citep{2021MNRAS.503.3975P} & \wrapc{Listed in variable/periodic and contact‐binary catalogues, but internal flags indicate it is neither EB nor variable.} \\
43.15890  & $+72.34585$ & \citep{2023AJ....166..193J} & \wrapc{Reported counts from LOFAR.} \\
84.77379  & $-01.48112$ & \citep{Dias} & \wrapc{Flagged with 89\% probability of membership in the Collinder 70 open cluster.} \\
69.11308  & $-21.21455$ & \citep{2019AJ....158...16S} & \wrapc{Flagged as RR Lyrae with RRab probability 0.0001.} \\
303.02484 & $+31.29017$ & \citep{2023MNRAS.521..354W} & \wrapc{Flagged as potential class II YSO; posterior probability $\simeq0$.} \\
\enddata
\end{deluxetable*}

\end{document}